\documentclass[aps,prx,twocolumn,superscriptaddress,longbibliography]{revtex4-2}

\usepackage{amssymb}
\usepackage{amsmath}
\usepackage{bm}
\usepackage{graphicx}
\usepackage{amsfonts}
\usepackage{braket}
\usepackage{subfigure}
\usepackage{hyperref,color}
\usepackage[version=3]{mhchem}

\begin{document}
\title{Two-fluid mobility model from coupled hydrodynamic equations for simulating laser-driven semiconductor switches}

\author{Qile Wu}
\email{qile\underline{ }wu@ucsb.edu.}
\affiliation{Physics Department, University of California, Santa Barbara, California 93106, USA}
\affiliation{Institute for Terahertz Science and Technology, University of California, Santa Barbara, California 93106, USA}

\author{Anton\'in Sojka}
\affiliation{Physics Department, University of California, Santa Barbara, California 93106, USA}
\affiliation{Institute for Terahertz Science and Technology, University of California, Santa Barbara, California 93106, USA}

\author{Brad D. Price}
\affiliation{Physics Department, University of California, Santa Barbara, California 93106, USA}
\affiliation{Institute for Terahertz Science and Technology, University of California, Santa Barbara, California 93106, USA}

\author{Nikolay I. Agladze}
\affiliation{Physics Department, University of California, Santa Barbara, California 93106, USA}
\affiliation{Institute for Terahertz Science and Technology, University of California, Santa Barbara, California 93106, USA}

\author{Anup Yadav}
\affiliation{School of Engineering, University of Warwick, Coventry, CV4 7AL, UK}

\author{Sophie L. Pain}
\affiliation{School of Engineering, University of Warwick, Coventry, CV4 7AL, UK}

\author{John D. Murphy}
\affiliation{School of Engineering, University of Warwick, Coventry, CV4 7AL, UK}

\author{Tim Niewelt}
\affiliation{School of Engineering, University of Warwick, Coventry, CV4 7AL, UK}
\affiliation{Fraunhofer Institute for Solar Energy Systems ISE, Heidenhofstraße 2, 79110 Freiburg, Germany}
\affiliation{Chair for Photovoltaic Energy Conversion, Department of Sustainable Systems Engineering, University of Freiburg, Emmy-Noether-Straße 2, 79110 Freiburg, Germany}

\author{Mark S. Sherwin}
\affiliation{Physics Department, University of California, Santa Barbara, California 93106, USA}
\affiliation{Institute for Terahertz Science and Technology, University of California, Santa Barbara, California 93106, USA}

\date{\today}

\begin{abstract}
We introduce a two-fluid mobility model incorporating fundamental aspects of electron-hole (e-h) scattering such as momentum conservation for simulating laser-driven semiconductor switches (LDSSs). Compared to previous works that use Matthiessen's rule, the two-fluid mobility model predicts distinct AC responses of e-h plasmas in semiconductors. Based on the two-fluid mobility model, we develop a theory with very few adjustable parameters for simulating the switching performance of LDSSs based on high-purity indirect-gap semiconductors such as silicon (Si). As a prototypical application, we successfully reproduce experimentally measured reflectance at around 320 GHz in a laser-driven Si switch. By injecting e-h plasmas with densities up to $10^{20}\,\rm cm^{-3}$, we reveal the importance of carrier-screening effects in e-h scattering and Auger recombination for carrier densities above the critical carrier density for exciton-plasma Mott transition. Our results also suggest a way to characterize the intrinsic momentum-relaxation mechanism, e-h scattering, and the intrinsic e-h recombination mechanism in indirect-gap semiconductors, Auger recombination. We reassess the ambipolar Auger coefficient of high-purity Si with high injection levels of e-h plasmas up to $10^{20}\,\rm cm^{-3}$ and find a minimal value of $1.8\times10^{-41}\,{\rm cm^6/ns}$. The value is more than one order of magnitude smaller than the ambipolar Auger coefficient widely used for simulating LDSSs, $3.8\times10^{-40}\,{\rm cm^6/ns}$, which was deduced from minority-carrier lifetime in highly doped silicon more than four decades ago.
\end{abstract}

\maketitle

Laser-driven semiconductor switches (LDSSs), originally introduced for generating short infrared pulses~\cite{alcock1975fast,jamison1978generation,corkum1978high,elezzabi1994generation,elezzabi1995600}, have been extensively studied for modulating terahertz (THz) radiation (0.1 to 10 THz)~\cite{salzmann1983subnanosecond,alius1991amplitude,vogel1992high,hegmann1996generation,nozokido1997modulation,doty2004wavelength,fekete2005active,picard2019laser,kutsaev2019nanosecond,denisov2021formation,denisov2021formation,li2023measurement,price2024compact}. Because of the lack of amplifiers, LDSSs are especially useful for shaping high-power THz sources ($>1$\,W), which have important applications in many fields such as plasma heating and diagnostics for nuclear fusion~\cite{paoloni2016thz,kariya2017development}, high-speed wireless communication~\cite{koenig2013wireless,akkacs2019terahertz}, high-resolution and nondestructive radar imaging~\cite{cooper2011thz,zhang2020three}, particle acceleration~\cite{nanni2015terahertz,zhang2020cascaded}, and manipulation of quantum states~\cite{salen2019matter,costello2021reconstruction}. The combination of LDSSs and optical cavities has also resulted in a way of generating high-power THz radiation through the cavity dumping technique~\cite{kaminski1990far,de1990generation,burghoorn1992generation,wilson1993high,genoud2022generation}.

In a LDSS, a semiconductor wafer is switched between a transmissive ``off'' state and a reflective ``on'' state. The switching is achieved via illumination by a pump laser, which generates an electron-hole (e-h) plasma in the material. Before the pump pulse arrives, a low-frequency electromagnetic wave that cannot excite either electron-hole pairs or phonons is almost fully transmitted through the wafer and the switch is ``off''. The switch is ``on'' when the pump pulse induces an e-h plasma of sufficiently high density in the semiconductor, which then behaves like a metal and strongly reflects the low-frequency radiation. After the pump pulse, the switch recovers back to the ``off'' state when the high-density plasma dissipates due to diffusion and e-h recombination. In the regime of linear response to the low-frequency radiation, understanding the electron and hole mobilities in e-h plasmas is the key to predicting the switching performance, as can be seen from the dielectric function~\cite{ashcroft1976solid},
\begin{align}
\varepsilon (\omega)
=
\varepsilon_{\rm L}
+
i\frac{\sigma(\omega)}{\varepsilon_0\omega}
,
\label{EQ:dielectric_func}
\end{align}
where $\varepsilon_{\rm L}$ is the contribution from the lattice vibration, $\varepsilon_0$ is the vacuum permittivity, $\omega$ is the angular frequency of the low-frequency radiation, and $\sigma(\omega)=Ne[\mu_{\rm e}(\omega)+\mu_{\rm h}(\omega)]$ is the total e-h conductivity defined by the e-h plasma density $N$, the elementary charge $e$, and the mobilities of the electrons and holes, $\mu_{\rm e}$ and $\mu_{\rm h}$. In previous studies, the mobilities of the electrons and holes in an e-h plasma are usually derived from the Drude model as~\cite{vogel1992high,nozokido1997modulation,picard2019laser,li2023measurement,schaub2021laser}
\begin{align}
\mu_{\rm e(h)}(\omega)
=
\frac
{e}{m_{\rm e(h)}}
\frac
{\tau_{\rm e(h)}}
{1-i\omega\tau_{\rm e(h)}},
\label{EQ:old_mobility_model}
\end{align}
where $m_{\rm e}$ and $m_{\rm h}$ are the conductivity effective masses~\cite{spitzer1957determination} of the electrons and holes, and the associated momentum relaxation times $\tau_{\rm e}$ and $\tau_{\rm h}$ are treated as two unrelated constants independent of the e-h plasma density. In fact, more than fifty years ago, it was experimentally shown that, when the electron and hole densities increase from $10^{14}$ to $10^{18}$ ${\rm cm}^{-3}$, the total mobilities of electrons and holes in undoped silicon (Si) could decrease drastically~\cite{dannhauser1972abhangigkeit,krausse1972abhangigkeit}. This phenomenon was quantitatively explained by taking account of e-h scattering as a dominant mechanism for momentum relaxation~\cite{dannhauser1972abhangigkeit,krausse1972abhangigkeit,fletcher1957high}. Since the density of e-h plasmas can easily go above $10^{17}$ ${\rm cm}^{-3}$ in a LDSS~\cite{vogel1992high,li2023measurement}, we expect that the dependence of the electron and hole mobilities on the e-h plasma density should be important in determining the switching performance. In 1999, T. E. Wilson made an attempt to include the carrier-density dependences of the momentum relaxation times, $\tau_{\rm e}$ and $\tau_{\rm h}$ in Eq.~(\ref{EQ:old_mobility_model}), in the simulation of LDSSs, by using ambipolar diffusivity data for doped Si~\cite{wilson1999modeling}. However, he ended up assuming that the electron and hole mobilities are both equal to the ambipolar mobility for the e-h plasma density above $10^{14}$ ${\rm cm}^{-3}$, because the electron and hole mobilities cannot be separately extracted from ambipolar diffusivity data, unless $\tau_{\rm e}$ and $\tau_{\rm h}$ have a well-defined relation, as in the two-fluid mobility model discussed below (see Appendix~\ref{APP:diffusion_coef} for a discussion).

For simulating the carrier-density dependence of the electron and hole mobilities in Si, in 1992, D. B. M. Klaassen came up with a physics-based analytical model~\cite{klaassen1992unified}, which has been used in recent reassessments of Auger recombination in crystalline Si~\cite{niewelt2022reassessment,black2022quantification}. In Klaassen's model, Matthiessen's rule is assumed for summing up contributions from various momentum-relaxation mechanisms including e-h scattering. We notice that, distinguished from other types of scattering such as ionized-impurity scattering, e-h scattering happens only when an electron and a hole have different velocities. Including this special feature of e-h scattering in the mobility model can lead to breakdown of Matthiessen's rule, and momentum conservation in e-h scattering imposes further constraints for modeling the electron and hole mobilities, as shown in a study of minority-carrier transport in GaAs quantum wells through two coupled hydrodynamic equations~\cite{hopfel1988electron}. As e-h scattering is important in the momentum relaxation of high-density e-h plasmas in typical LDSSs, a mobility model incorporating the aforementioned fundamental aspects of e-h scattering is desired.

In this paper, we derive a two-fluid mobility model for simulating LDSSs based on the coupled hydrodynamic equations introduced in Ref.~\cite{hopfel1988electron}. The two-fluid mobility model, in which the electron and hole mobilities are parametrized through the same set of parameters, leads us to a theory with very few adjustable parameters for simulating LDSSs based on high-purity indirect-gap semiconductors such as Si. As a prototypical application, we demonstrate how our theory can be used to explain the experimentally measured reflectance at around 320 GHz in a laser-driven Si switch,  in which the Si wafer is of sufficiently high purity and the surfaces are sufficiently well-passivated that the dominant e-h recombination mechanism is Auger recombination. The choice of such a high-purity sample comes with its own significance. Even for Si, the most frequently used semiconductor material, recent years have seen significant improvement in sample quality, and reassessment of the intrinsic properties of Si is required for more refined design of semiconductor devices~\cite{niewelt2022reassessment,black2022quantification}. Following the progress in surface passivation technology and wafer pretreatment, an extraordinarily high minority-carrier lifetime of 0.5\,s in n-type Si has been reported~\cite{steinhauser2021extraordinarily}. By investigating a large number of Si wafers processed by state-of-the-art surface passivation, the Auger recombination process in Si with carrier-injection levels ranging from $10^{14}$ to $10^{17}$ ${\rm cm}^{-3}$ has been reassessed in 2022 through photoconductance-decay and modulated luminescence experiments~\cite{niewelt2022reassessment}. Here, by fitting the simulated reflectance to the experimental data, we reassess the ambipolar Auger recombination in Si with high injection levels up to $10^{20}$ ${\rm cm}^{-3}$ and find a minimal ambipolar Auger coefficient of $1.8\times10^{-41}\,{\rm cm^6/ns}$. The value is more than one order of magnitude smaller than the ambipolar Auger coefficient widely used for simulating LDSSs, $3.8\times10^{-40}\,{\rm cm^6/ns}$, which was deduced from minority-carrier lifetime in highly doped silicon in 1977~\cite{dziewior1977auger}. In our approach, two intrinsic processes of indirect-gap semiconductors, which include the intrinsic momentum-relaxation process, e-h scattering, and the intrinsic e-h recombination process, Auger recombination, are characterized at the same time.

\begin{figure}
	\includegraphics[width=0.47\textwidth]{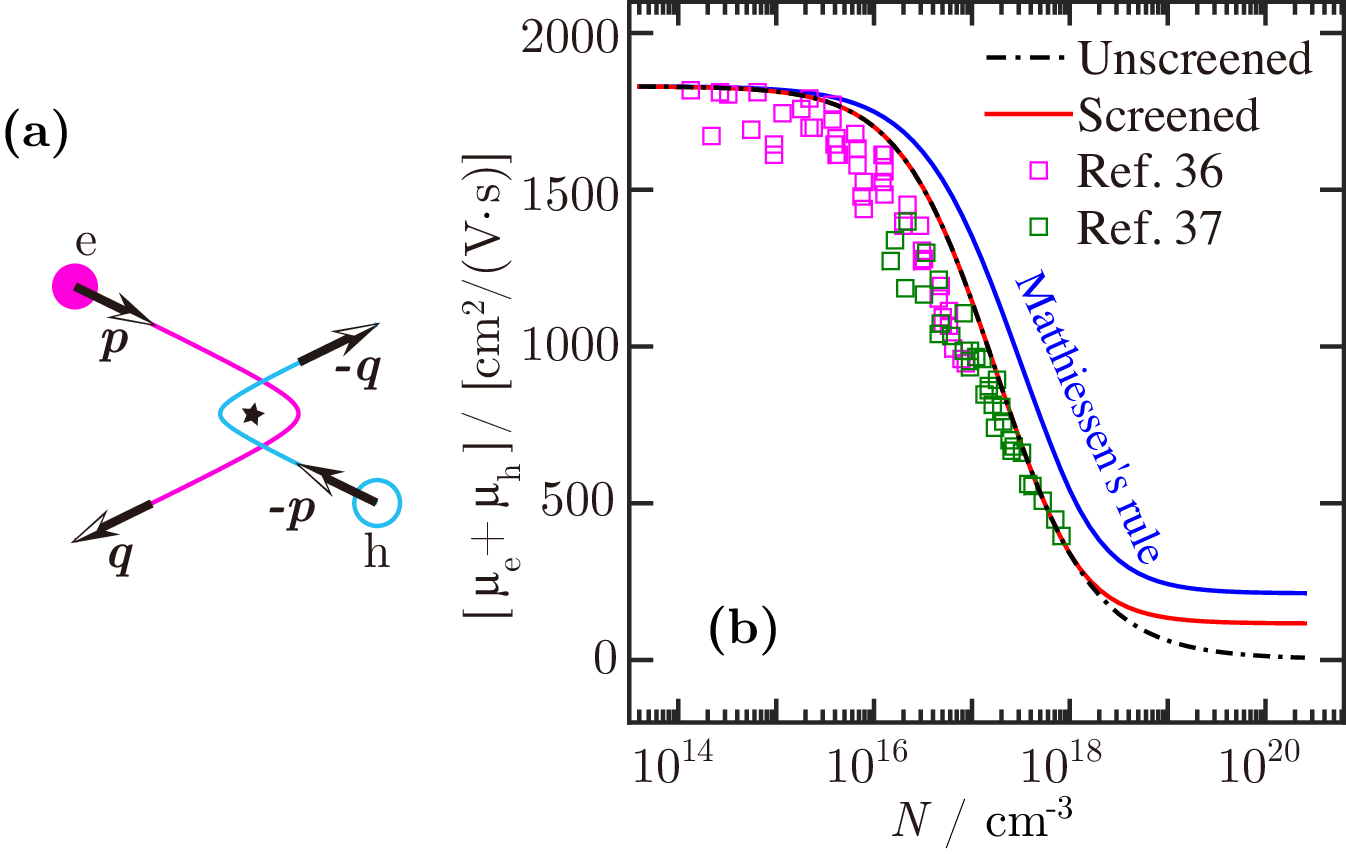}
	\caption{Mobility model incorporating momentum conservation in electron-hole scattering. (a) Classical electron-hole (e-h) scattering in the center-of-mass reference frame. An electron (magenta ball) with momentum ${\bf p}$ and a hole (cyan circle) with momentum $-{\bf p}$ scatter with each other and end up having different momenta, ${\bf q}$ and $-{\bf q}$. The black star represents their center of mass. (b) Comparison between experimentally measured and theoretically calculated sums of electron and hole mobilities in DC electric fields for undoped silicon (Si) at 300 K. The magenta (dark-green) squares represent the data from Ref.~\cite{dannhauser1972abhangigkeit} (Ref.~\cite{krausse1972abhangigkeit}). The black dash-dotted (red solid) line represents the result calculated from Eqs.~(\ref{EQ:Mobility_sum}) and~(\ref{EQ:Mobility_eh}) without (with) corrections from carrier-screening effects. The blue curve shows the mobility sum calculated by using Matthiessen's rule.}
	\label{FIG:mobility_model}
\end{figure}

Following Ref.~\cite{hopfel1988electron}, we describe the momentum relaxation of the e-h plasmas in an oscillating electric field ${\bf E}\exp({-i\omega t})$ with two coupled hydrodynamic equations:
\begin{align}
\frac{d{\bf v}_{\rm e}}{dt}
=
-
\frac{{\bf v}_{\rm e}}{\tau'_{\rm e}}
-
\frac{{\bf v}_{\rm e}-{\bf v}_{\rm h}}{\tau_{\rm e-h}}
-\frac{e\bf E}{m_{\rm e}}\exp({-i\omega t}),
\label{EQ:Momentum_relaxation_electron}
\\
\frac{d{\bf v}_{\rm h}}{dt}
=
-
\frac{{\bf v}_{\rm h}}{\tau'_{\rm h}}
-
\frac{{\bf v}_{\rm h}-{\bf v}_{\rm e}}{\tau_{\rm h-e}}
+\frac{e\bf E}{m_{\rm h}}\exp({-i\omega t}),
\label{EQ:Momentum_relaxation_hole}
\end{align}
where ${\bf v}_{\rm e(h)}$ is the mean velocity of the electrons (holes), $\tau_{\rm e-h}$ and $\tau_{\rm h-e}$ are momentum relaxation times associated with e-h scattering, and $\tau'_{\rm e(h)}$ is the momentum relaxation time associated with other scattering mechanisms for electrons (holes). In this paper, we use the convention that the carrier velocities and electromagnetic fields are written in complex forms, whose real parts are taken as the corresponding physical quantities. In an e-h scattering event, the center-of-mass momentum remains the same, while the relative momentum changes direction (Fig.~\ref{FIG:mobility_model} (a)). The second term on the right-hand side of Eq.~(\ref{EQ:Momentum_relaxation_electron}) or (\ref{EQ:Momentum_relaxation_hole}) is consistent with the fact that momentum relaxation does not occur in e-h scattering if an electron and a hole have the same velocity. In addition, momentum conservation in e-h scattering requires that
\begin{align}
-
m_{\rm e}\frac{{\bf v}_{\rm e}-{\bf v}_{\rm h}}{\tau_{\rm e-h}}
-
m_{\rm h}\frac{{\bf v}_{\rm h}-{\bf v}_{\rm e}}{\tau_{\rm h-e}}
=0,
\end{align}
based on which we define $\mu_{\rm eh}\equiv{e\tau_{\rm e-h}}/{ m_{\rm e}}={e \tau_{\rm h-e}}/{m_{\rm h}}$ to characterize the momentum relaxation associated with e-h scattering. Using the ansatz ${\bf v}_{\rm e(h)}=\mu_{\rm e(h)}(\omega){\bf E}\exp({-i\omega t})$ in Eqs.~(\ref{EQ:Momentum_relaxation_electron}) and (\ref{EQ:Momentum_relaxation_hole}), we solve the resulting linear equations and obtain the electron and hole mobilities, $\mu_{\rm e}(\omega)$ and $\mu_{\rm h}(\omega)$, in the following forms:
\begin{align}
\mu_{\rm e}(\omega)
=
[\mu^{-1}_{\rm e,0}-i\frac{m_{\rm e}\omega}{e}
+
(1+\frac{\mu^{-1}_{\rm e,0}-i\frac{m_{\rm e}\omega}{e}}{\mu^{-1}_{h,0}-i\frac{m_{\rm h}\omega}{e}})\mu^{-1}_{\rm eh}]^{-1},
\label{EQ:Mobility_electron}
\\
\mu_{\rm h}(\omega)
=
[\mu^{-1}_{\rm h,0}-i\frac{m_{\rm h}\omega}{e}
+
(1+\frac{\mu^{-1}_{\rm h,0}-i\frac{m_{\rm h}\omega}{e}}{\mu^{-1}_{e,0}-i\frac{m_{\rm e}\omega}{e}})\mu^{-1}_{\rm eh}]^{-1},
\label{EQ:Mobility_hole}
\end{align}
where $\mu_{\rm e(h),0}\equiv{e \tau'_{\rm e(h)}}/{m_{\rm e(h)}}$ is the electron (hole) mobility in DC electric fields in the absence of e-h scattering. Equations~(\ref{EQ:Mobility_electron}) and~(\ref{EQ:Mobility_hole}) both reduce to the form of Eq.~(\ref{EQ:old_mobility_model}) when e-h scattering is ignored. Compared to Eq.~(\ref{EQ:old_mobility_model}), which is used in previous work for simulating LDSSs, the two-fluid mobility model predicts a different frequency dependence in the dielectric function, and thus distinct AC responses. An interesting feature of the two-fluid mobility model is that the electron and hole mobilities are parametrized by the same set of parameters, $\mu_{\rm e,0}$, $\mu_{\rm h,0}$, and $\mu_{\rm eh}$. In the following, we apply the two-fluid mobility model to the simulation of a LDSS, in which the reflectance variation results from AC responses of optically injected e-h plasmas to sub-THz fields ($\omega=2\pi\times314.4\,{\rm GHz}$).

To demonstrate how the two-fluid mobility model can be applied in simulating LDSSs, we consider a widely used material---Si at 300\,K. We focus on high-purity Si samples, in which the dominant momentum-relaxation mechanism is either carrier-phonon scattering or e-h scattering. We first check if the mobility model is consistent with existing experimental data from Ref.~\cite{dannhauser1972abhangigkeit} and Ref.~\cite{krausse1972abhangigkeit} for the sum of electron and hole mobilities in DC electric fields (Fig.~\ref{FIG:mobility_model} (b)), and at the same time determine the parameters, $\mu_{\rm e,0}$, $\mu_{\rm h,0}$, and $\mu_{\rm eh}$, in Eqs.~(\ref{EQ:Mobility_electron}) and~(\ref{EQ:Mobility_hole}). In DC electric fields ($\omega=0$), the sum of the electron and hole mobilities has the simple form,
\begin{align}
\mu_{\rm e}(0)
+
\mu_{\rm h}(0)
=
[
({\mu_{\rm e,0}+\mu_{\rm h,0}})^{-1}
+
\mu^{-1}_{\rm eh}]^{-1}.
\label{EQ:Mobility_sum}
\end{align}
We take $\mu_{\rm e,0}=1350$ ${\rm cm^2/(V\cdot s)}$ and $\mu_{\rm h,0}=480$ ${\rm cm^2/(V\cdot s)}$ as constants from mobility measurements on Si single crystals~\cite{ludwig1956drift}, and assume that these mobility values are mainly contributed from carrier-phonon scattering such that they can be used for undoped Si. Following the original explanation of the data in Ref.~\cite{dannhauser1972abhangigkeit} and Ref.~\cite{krausse1972abhangigkeit}, we try the formula describing classical mutual diffusion of two gases~\cite{chapman1990mathematical},
\begin{align}
\mu_{\rm eh}
=
\frac
{3e (k_{\rm B}T)^{3/2}}
{
N\alpha^2\sqrt{8\pi m_{\rm eh}}
\ln[1+\frac{(4k_{\rm B}T)^2}{\alpha^2N^{2/3}}]
},
\label{EQ:Mobility_eh}
\end{align}
where $k_{\rm B}$ is the Boltzmann constant, $T=300$\,K is the carrier temperature assumed to be the same as the lattice temperature, $m_{\rm eh}=m_{\rm e}m_{\rm h}/(m_{\rm e}+m_{\rm h})$ is the reduced mass of the e-h pairs, and $\alpha=e^2/(4\pi\epsilon_0\epsilon_{\rm L,0})$ defines the Coulomb interaction strength with $\epsilon_{\rm L,0}=3.4175^2$ being the lattice contribution to the static dielectric constant~\cite{dai2004terahertz}. Equation~(\ref{EQ:Mobility_eh}) can be derived by applying the Conwell-Weisskof approach~\cite{conwell1950theory} to e-h scattering and adding a constant correction factor to account for the effects of electron-electron scattering and hole-hole scattering~\cite{fletcher1957high}. The black dash-dotted line in Fig.~\ref{FIG:mobility_model} (b) shows the mobility sum $\mu_e+\mu_h$ calculated by using Eqs.~(\ref{EQ:Mobility_sum}) and~(\ref{EQ:Mobility_eh}). Here, we take $m_{\rm e}=0.2558m_0$ ($m_0$ is the electron rest mass), which is an average~\cite{spitzer1957determination} over six equivalent conduction-band minima with transverse and longitudinal effective masses, $m_{\rm t}=0.1905m_0$ and $m_{\rm l}=0.9163m_0$~\cite{hensel1965cyclotron}, and take $m_{\rm h}=0.3640m_0$, which is an average~\cite{spitzer1957determination} over the two highest valence bands with their conductivity effective masses averaged over all orientations by using the Luttinger parameters, $\gamma_1=4.285$,
$\gamma_2=0.339$, and 
$\gamma_3=1.446$~\cite{hensel1977private}. Surprisingly, the carrier-density dependence in the mobility-sum data is well reproduced through Eq.~(\ref{EQ:Mobility_eh}) without any empirically adjustable parameter. Note that the original explanation of the data
assumed Matthiessen's rule~\cite{dannhauser1972abhangigkeit,krausse1972abhangigkeit,fletcher1957high}, i.e., $1/\mu_{\rm e(h)}=1/\mu_{\rm e(h),0}+1/\mu_{\rm eh}$, which would lead to an overestimate of the mobility sum if the above more up-to-date parameters were used (blue curve in Fig.~\ref{FIG:mobility_model} (b)), and Klaassen's model could match with the data only if the sample was assumed to be heated up by 200\,K~\cite{klaassen1992unified}.

\begin{figure}
	\includegraphics[width=0.47\textwidth]{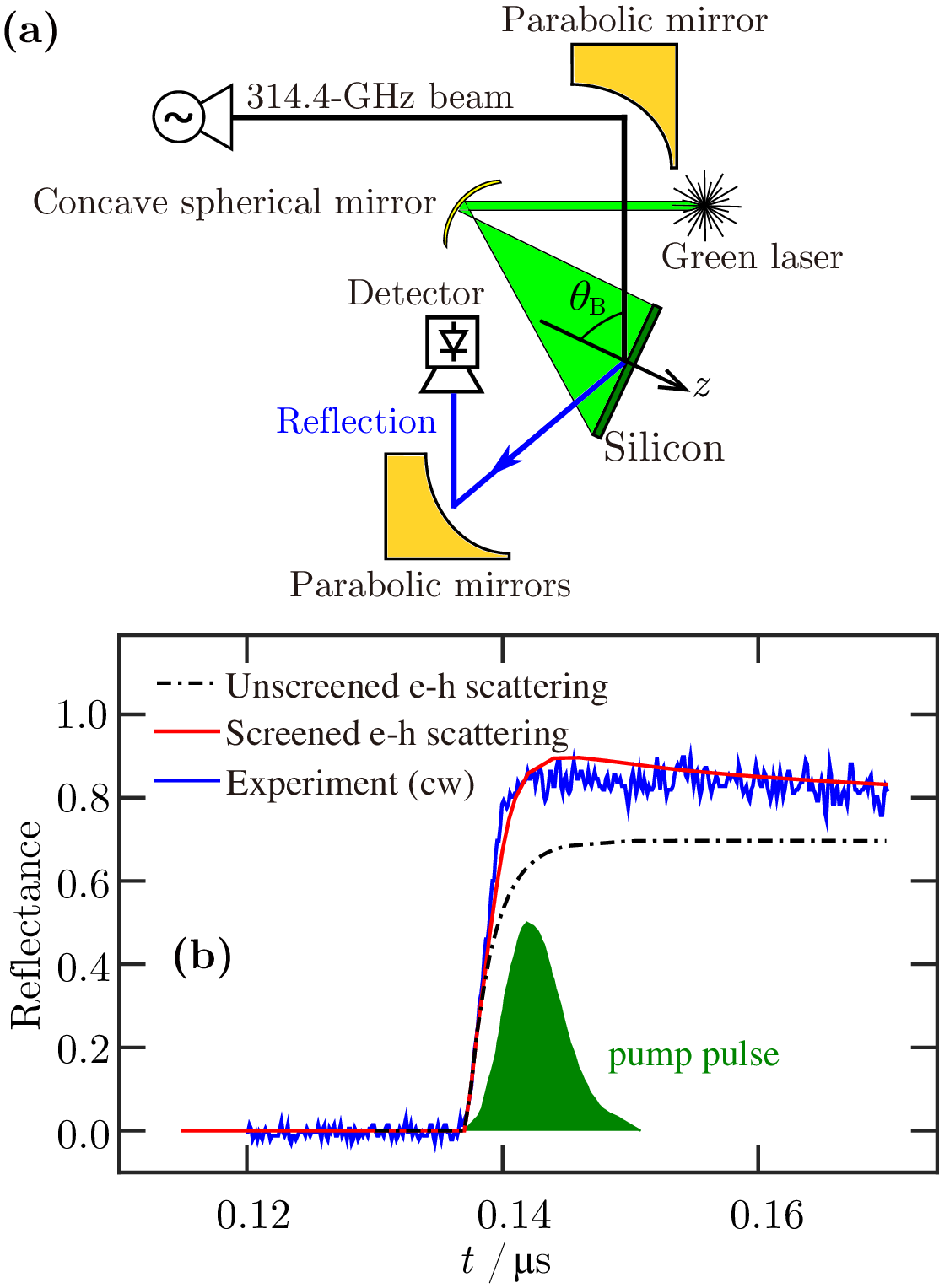}
	\caption{Reflectance measurement in a laser-driven silicon (Si) switch and Si reflectance around the laser excitation. (a) Schematic diagram of the experimental setup. Sub-THz waves with a frequency of 314.4\,GHz were directed by a parabolic mirror onto a Si wafer at Brewster's angle $\theta_{\rm B}$. The reflection signals were passed by several parabolic mirrors to a quasi-optical detector. A green-laser beam was expanded by a concave spherical mirror and incident on the Si-wafer surface at an angle of about $45^{\circ}$ to generate e-h plasmas. (b) Comparison between measured and simulated reflectances for the continuous-wave (cw) experiment. The blue curve represents the reflectance measured with a continuous sub-THz wave. The black dash-dotted line represents the simulated results without including carrier-screening effects in e-h scattering and with the assumption of no e-h recombination. The red solid line represents the simulated results with corrections from carrier-screening effects in e-h scattering and a constant Auger coefficient $r_{\rm Auger}=1.8\times10^{-41}\,{\rm cm^6/ns}$. The sums of electron and hole mobilities corresponding to the black dash-dotted and red solid lines are shown as black dash-dotted and red solid lines, respectively, in Fig.~\ref{FIG:mobility_model} (b). The dark-green shaded area shows the temporal profile of the green-laser pulse.}
	\label{FIG:simulation_refl_cw}
\end{figure}
Based on the consistency between the DC-field mobility data and the two-fluid mobility model with the e-h scattering described by Eq.~(\ref{EQ:Mobility_eh}), we proceed to investigate whether the two-fluid mobility model can be used for simulating AC responses of Si. Specifically, we examine here the reflectance of Si for sub-THz waves in a laser-driven Si switch shown in Fig.~\ref{FIG:simulation_refl_cw} (a). A 5-ns, 55.7-mJ green-laser pulse (see dark-green shaded area in Fig.~\ref{FIG:simulation_refl_cw} (b) for the measured power profile) with a central wavelength of 532.263\,nm (DPS-532-BS-D-50mJ, Changchun New Industries Optoelectronics Tech. Co., Ltd., CN) was incident on the Si-wafer surface at an angle of about $45^{\circ}$ to generate an e-h plasma. The electric fields in the green-laser pulses were set to be in the plane of incidence with respect to the Si-wafer surface, i.e., p-polarization, to maximize their transmission into the Si. To fully cover the sub-THz beam spot on the Si wafer, the green-laser beam was expanded by a concave spherical mirror into an elliptical shape. Sub-THz waves with a frequency of 314.4\,GHz were generated by a microwave generator (SynthHD Mini, Windfreak Technologies, LLC USA) and an amplifier/multiplier chain (Custom made AMC, Virginia Diodes, Inc., USA) and then directed by a parabolic mirror onto a 725-${\rm \mu m}$-thick n-type Si wafer (resistivity in the range of 500 to 1200 $\rm \Omega\cdot cm$) at Brewster's angle $\theta_{\rm B}=\arctan[\sqrt{{\rm Re}(\varepsilon_{\rm L})}/n_{\rm Air}]\approx73.7^{\circ}$, where $\varepsilon_{\rm L}=(3.4175+0.00002i)^2$ is the dielectric constant of Si at 314.4\,GHz (see Appendix~\ref{APP:lattice_dielectric} for a discussion), and $n_{\rm Air}=1.0003$ is the refractive index of the air~\cite{ciddor1996refractive}. The electric fields in the sub-THz waves were also set to be p-polarization, which resulted in a minimal reflectance before the green-laser pulse arrived. The reflection signals from the Si wafer were passed by several parabolic mirrors to a quasi-optical detector (3DL 12C LS2500 A2,ACST GmbH, DE). The Si wafer and parabolic mirrors mentioned here were all embedded in a recently developed compact module for THz-pulse slicing~\cite{price2024compact}.

\begin{figure}
	\includegraphics[width=0.47\textwidth]{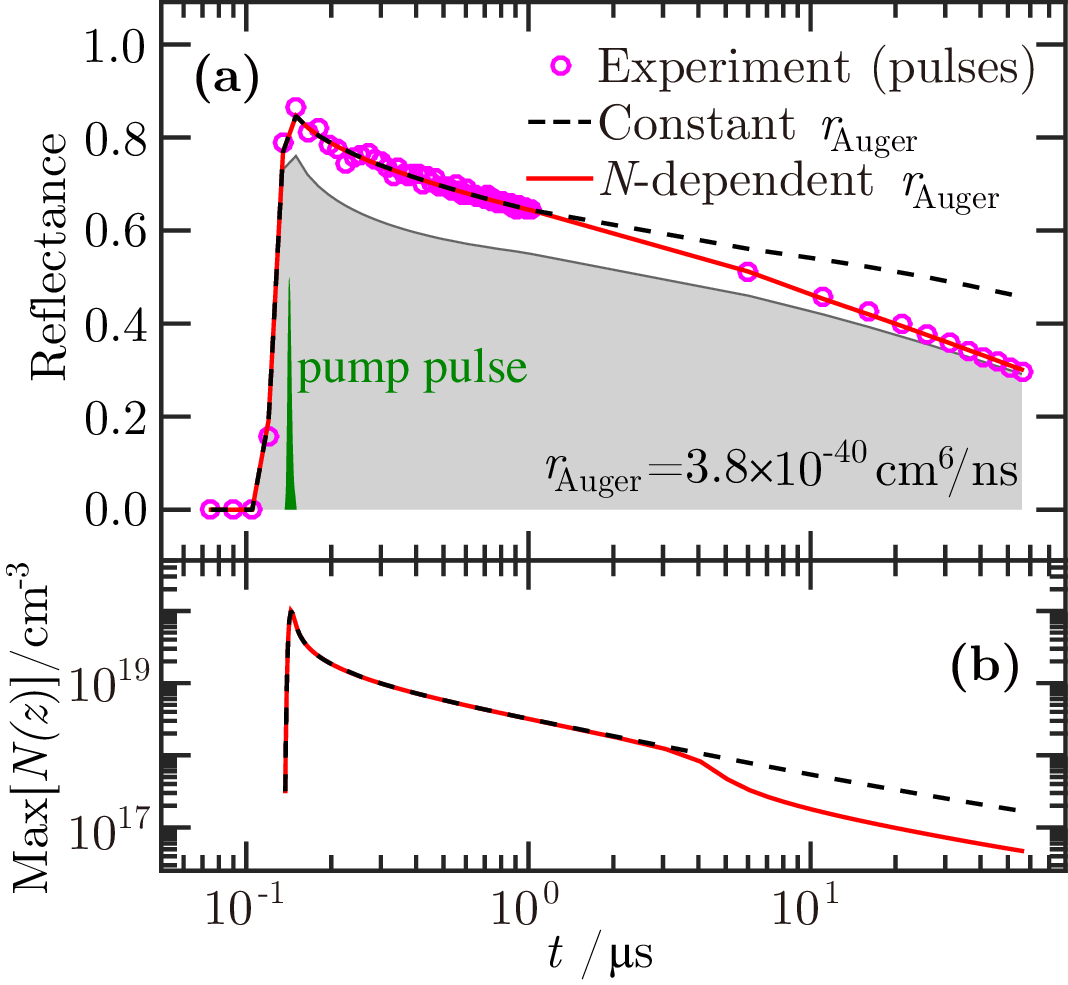}
	\caption{Silicon (Si) reflectance and carrier dynamics up to 60\,$\mu$s after the pump pulse. (a) Comparison between measured and simulated reflectance values for the pulse experiment. Each reflectance value is an average over  a 25-ns sub-THz pulse and is plotted at the time corresponding to the rising edge of the sub-THz pulse. The magenta circles represent the measured reflectance values. The red solid line represents the results simulated with the Auger coefficient $r_{\rm Auger}$ described by Eq.~(\ref{EQ:r_auger}). The black dashed line shows the results simulated with $r_{\rm Auger}=1.8\times10^{-41}\,{\rm cm^6/ns}$. The shaded area shows the results simulated by using $r_{\rm Auger}=3.8\times10^{-40}\,{\rm cm^6/ns}$, which is widely used for simulating laser-driven Si Switches~\cite{vogel1992high,nozokido1997modulation,picard2019laser,li2023measurement,schaub2021laser}. The dark-green shaded area shows the temporal profile of the green-laser pulse. (b) The maximal e-h plasma density $N$ in the Si wafer as a function of time. The red solid and black dashed lines corresponds to the red solid and black dashed lines in (a), respectively.}
	\label{FIG:simulation_refl_pulse}
\end{figure}

To resolve the rising edge in the temporal profile of the reflectance, we first used a continuous sub-THz wave and monitored the reflection signals through an oscilloscope (Lecroy WavePro 950) around the green-laser pulse. Because the quasi-optical detector is only sensitive to changes in the input power, the reflection signals from the continuous sub-THz wave differed from the actual reflectance by a constant factor, which could not be obtained by sending the continuous sub-THz wave directly into the detector. To normalize the reflection signals from the continuous-wave (cw) experiment, further reflectance measurements were done by employing a PIN diode-based switch between the microwave generator and the amplifier/multiplier chain to create 25-ns sub-THz pulses. In the pulse experiment, the reflectance values were averaged in 25-ns time windows by normalizing each reflection signal to a reference signal, which was measured by sending the corresponding 25-ns sub-THz pulse directly into the detector. The normalization of the reflection signals from the continuous sub-THz wave was achieved by making use of the connection between the cw and pulse experiments. The blue curve in Fig.~\ref{FIG:simulation_refl_cw} (b) shows the normalized reflection signal in the cw experiment (See Appendix~\ref{APP:reflectance_exp} for more details about the experiments and Appendix~\ref{APP:r_normalization} for details about the reflectance normalization). The readout of the quasi-optical detector is more reliable for signals shorter than 100\,ns because its internal amplifier has diminished responses at frequencies lower than 8\,MHz, so we focus on a 50-ns time window around the laser excitation for the reflection signals from the continuous sub-THz wave. The switching performance on a microsecond time scale was investigated through the pulse experiment, in which the averaged reflectance values were recorded as a function of the time at the rising edge of each 25-ns sub-THz pulse (magenta circles in Fig.~\ref{FIG:simulation_refl_pulse} (a)). The reflectance reached above 0.85 during the green-laser pulse, remained above 0.6 about 1\,$\mu$s after the pump pulse, and stayed above 0.3 50\,$\mu$s later.

To simulate the laser-driven Si switch, we calculate the e-h plasma density through the continuity equation,
\begin{align}
\frac{\partial N(z,t)}{\partial t}
=
\frac{\partial J}{\partial z}
+G(z,t)-R(N),
\label{EQ:continuity_n}
\end{align}
where $J=D(N)({\partial N}/{\partial z})$ is the ambipolar diffusion-current density with $D(N)$ being the coefficient of amipolar diffusion along the direction normal to the wafer surfaces ($z$-axis in Fig.~\ref{FIG:simulation_refl_cw} (a)), $G(z,t)=\eta_{\rm abs}(z,t)/E_{\rm gr}$ is the e-h generation rate determined by the energy-density loss rate $\eta_{\rm abs}(z,t)$ of the green-laser pulse in the absorptive dissipation in the Si wafer and the photon energy $E_{\rm gr}\approx2.33\,{\rm eV}$ associated with the central wavelength of the green laser, and $R(N)$ is the e-h recombination rate. The energy-density loss rate $\eta_{\rm abs}(z,t)$ is calculated by considering the propagation of a plane electromagnetic wave in an air-Si-air structure (see Appendix~\ref{APP:light_propagation} and Appendix~\ref{APP:green_light} for details). Here, we approximate the intensity of the green laser to be uniform across the sub-THz beam spot on the Si-wafer surface, and that all the green-laser photons absorbed by Si are converted to e-h pairs. The peak intensity of the green-laser pulse is determined to be about 4.15\,${\rm MW/cm^2}$ by fitting the rising edge of the reflectance at the beginning of the laser excitation (from 0.137 to 0.138\,$\mu$s in Fig.~\ref{FIG:simulation_refl_cw} (a)). As the green-laser pulse propagates in the Si wafer, its intensity is attenuated, resulting in inhomogeneity in the e-h plasma density $N$ along the $z$-axis. The $z$-dependence of the carrier density leads to diffusion of the electrons and holes. While the electrons diffuse faster than the holes for the same carrier concentration, e-h Coulomb attraction tends to diminish the velocity difference between the electrons and holes.
Under the assumption of local charge neutrality, the electrons and holes move together with the same ambipolar diffusion-current density $J$. The ambipolar diffusion coefficient $D(N)$, which is usually taken as a constant independent of the electron and hole mobilities~\cite{vogel1992high,nozokido1997modulation,picard2019laser,li2023measurement,schaub2021laser}, is determined here by the DC-field mobilities $\mu_e(0)$ and $\mu_h(0)$ as (see Appendix~\ref{APP:diffusion_coef} for a derivation)
\begin{align}
D(N)
=
\frac{2k_BT}{e}\frac{\mu_e(0)\mu_h(0)}{\mu_h(0)+\mu_e(0)}.
\end{align}
The electrons and holes are assumed to be separately in thermal equilibrium with the lattice and have the same temperature $T=300$ K. To simplify the analysis of e-h recombination, the Si wafer was chosen to be sufficiently pure to minimize the effects of Shockley-Read-Hall (SRH) recombination at bulk defects, and a 25-nm aluminum-oxide layer was deposited onto each of the Si-wafer surfaces to reduce the surface recombination velocity to an extremely low level of less than 1\,cm/s~\cite{agostinelli2006very,dingemans2010silicon,grant2024activation}. Before the reflectance measurements, the SRH recombination time was estimated to be about 30\,ms through photoconductance-decay measurements with a Sinton WCT-120 lifetime tester~\cite{sinton1996contactless} (see Appendix~\ref{APP:si_lifetime} for the data). Considering that the radiative recombination is relatively unimportant in undoped Si with high photocarrier concentrations~\cite{niewelt2022reassessment,black2022quantification}, we take Auger recombination as the only e-h recombination mechanism in the simulation and write
\begin{align}
R(N)= r_{\rm Auger}N^3,
\end{align}
with $r_{\rm Auger}$ being the ambipolar Auger coefficient. The carriers in the Si wafer before optical injection are ignored, since they do not have significant effects on the reflectance (see Fig.~\ref{FIG:simulation_refl_cw} (b) and Fig.~\ref{FIG:simulation_refl_pulse} (a)). The continuity equation, Eq.~(\ref{EQ:continuity_n}), is solved numerically by using the Crank-Nicolson method (see Appendix~\ref{APP:numerical_plasma} for details) to obtain the e-h plasma density $N(z,t)$, which, together with the two-fluid mobility model (Eqs.~(\ref{EQ:Mobility_electron}) and~(\ref{EQ:Mobility_hole})), determines the dielectric function through Eq.~(\ref{EQ:dielectric_func}). The reflectance of the optically excited Si wafer that has a $z$-dependent dielectric function is calculated by considering the propagation of a plane electromagnetic wave in a stratified medium~\cite{born2013principles} (see Appendix~\ref{APP:light_propagation} and Appendix~\ref{APP:r_subthz} for details). 

\begin{figure}
	\includegraphics[width=0.47\textwidth]{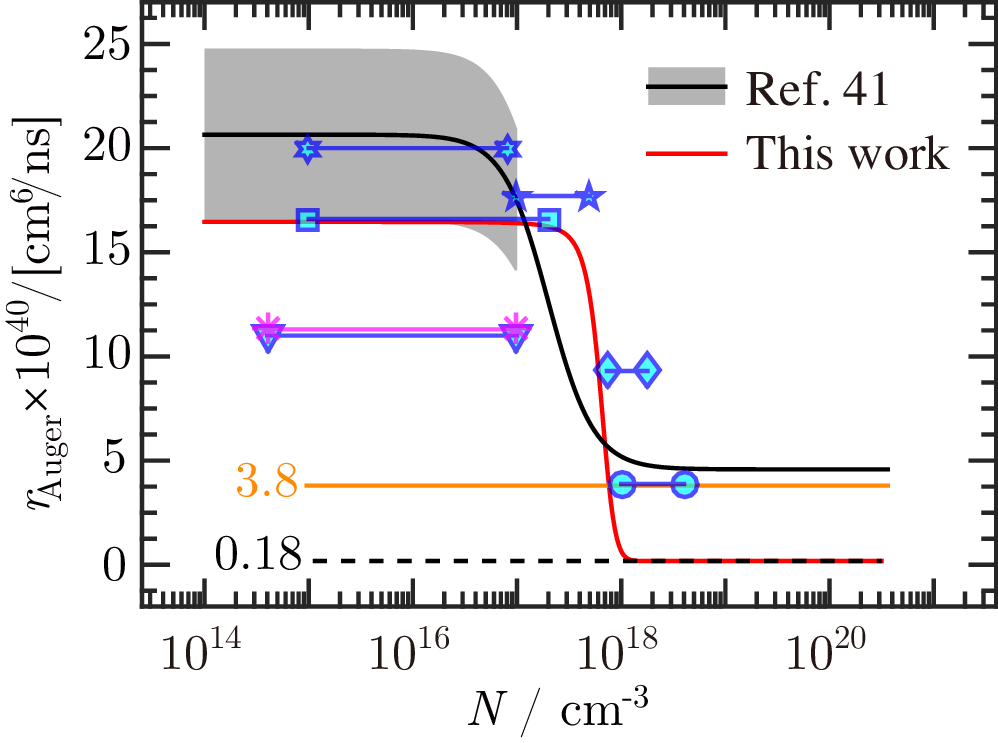}
	\caption{Ambipolar Auger coefficient $r_{\rm Auger}$ as a function of e-h plasma density $N$. The red solid line is described by Eq.~(\ref{EQ:r_auger}) and the black dashed line indicates the minimal value $r_{\rm min}=1.8\times10^{-41}\,{\rm cm^6/ns}$. The red solid and black dashed lines correspond to the red solid and black dashed lines in Fig.~\ref{FIG:simulation_refl_pulse}, respectively. The black solid line represents the results from an empirical formula used in Ref.~\cite{niewelt2022reassessment}, in which the carrier-injection levels range from $10^{14}$ to $10^{17}\,{\rm cm^{-3}}$ and the relative deviations of the extracted Auger recombination rates from the empirical formula are mostly within 20\% (shaded area). The scattered markers represent data from the literature: downward-pointing triangles (Ref.~\cite{pang1991record}), asterisks (Ref.~\cite{pang1993new}), squares (Ref.~\cite{sinton1987recombination}), pentagrams (Ref.~\cite{misiakos1990carrier}), hexagrams (Ref.~\cite{yablonovitch1986auger}), diamonds (Ref.~\cite{forget1992auger}), circles (Ref.~\cite{svantesson1979determination}). Each horizontal line connecting two identical markers indicate the corresponding carrier-density range. The levels of carrier injection for the data from Ref.~\cite{forget1992auger} are estimated by using Eq.\,(5) in the reference. The orange line indicates the ambipolar Auger coefficient widely used for simulating laser-driven Si switches, $3.8\times10^{-40}\,{\rm cm^6/ns}$~\cite{vogel1992high,nozokido1997modulation,picard2019laser,li2023measurement,schaub2021laser}, which was deduced from luminescence-decay measurements in highly doped silicons in Ref.~\cite{dziewior1977auger}.}
	\label{FIG:auger_coefficient}
\end{figure}

With the parameter $\mu_{\rm eh}$ in the two-fluid mobilty model described by using Eq.~(\ref{EQ:Mobility_eh}), which contains no adjustable parameters, the simulated reflectance profile (black dash-dotted line in Fig.~\ref{FIG:simulation_refl_cw} (b)) for the cw experiment lies about 0.1 below the measured curve (blue curve), even when no e-h recombination is present. To resolve this disagreement, we consider that, around the laser excitation, the maximal e-h plasma density generated by the green laser is on the order of $10^{20}\,{\rm cm^{-3}}$ (see Appendix~\ref{APP:green_light} for an estimate and Fig.~\ref{FIG:simulation_refl_pulse} (b) for a result from numerical simulation), well above the critical carrier density ($N_{\rm Mott}\approx 5\times 10^{17}\,{\rm cm^{-3}}$) for exciton-plasma Mott transition in Si~\cite{norris1982exciton}. For such a high e-h plasma density, the e-h scattering model, Eq.~(\ref{EQ:Mobility_eh}), which is derived by using unscreened e-h Coulomb interaction, could have underestimated the electron and hole mobilities. To maintain the consistency between the two-fluid mobility model and the DC-field mobility data for e-h plasma density $N$ below $10^{18}\,{\rm cm^{-3}}$ (squares in Fig.~\ref{FIG:mobility_model}), we modify the parametrization of the two-fluid mobility model by keeping Eq.~(\ref{EQ:Mobility_eh}) for $N\le10^{18}\,{\rm cm^{-3}}$ and lifting up the electron and hole mobilities by redefining the values of $\mu_{\rm eh}$ for $N>10^{18}\,{\rm cm^{-3}}$ through a simple function, 
\begin{align}
\mu_{\rm eh}(N)=e^{A-B\log_{10}N}+\mu_{\rm eh,\infty}, 
\label{EQ:mueh_modified}
\end{align}
which exponentially decreases to a constant $\mu_{\rm eh,\infty}$ with respect to $\log_{10} N$. Here, $A=53.266$ and $B=2.6438$ are constants set by matching the function value and slope with respect to $\log_{10} N$ given by Eq.~(\ref{EQ:mueh_modified}) to those given by Eq.~(\ref{EQ:Mobility_eh}) at $N=10^{17.911}\,{\rm cm^{-3}}$, which corresponds to the lowest mobility-sum data point from Ref.~\cite{krausse1972abhangigkeit} (Fig.~\ref{FIG:mobility_model} (b)).
With the mobility constant $\mu_{\rm eh,\infty}$ and the Auger coefficient $r_{\rm Auger}$ treated as the only two adjustable parameters, we find that the reflectance data from both the cw and pulse experiments can be well reproduced up to 1\,$\mu$s after the pump pulse. To obtain an optimal set of parameters, we iteratively tune $\mu_{\rm eh,\infty}$ to minimize the deviation of the simulated reflectance from that measured in the cw experiment (blue curve in Fig.~\ref{FIG:simulation_refl_cw} (b)) and tune $r_{\rm Auger}$ to minimize the deviation of the simulated reflectance values from those measured with sub-THz pulses delayed by less than 3\,$\mu$s with respect to the pump pulse (magenta circles in Fig.~\ref{FIG:simulation_refl_pulse} (a)). The reflectance values simulated with the optimal set of parameters, $\mu_{\rm eh,\infty}=124\,\rm cm^2/(V\cdot s)$ and $r_{\rm Auger}=1.8\times10^{-41}\,{\rm cm^6/ns}\equiv r_{\rm min}$ (black dashed line in Fig.~\ref{FIG:auger_coefficient}), are shown as red solid line in Fig.~\ref{FIG:simulation_refl_cw} (b) for the cw experiment and black dashed line in Fig.~\ref{FIG:simulation_refl_pulse} (a) for the pulse experiment, respectively. 

To explain the reflectance values measured with sub-THz pulses delayed by more than 3\,$\mu$s with respect to the pump pulse (magenta circles in Fig.~\ref{FIG:simulation_refl_pulse} (a)), we notice that, in the simulation with a constant Auger coefficient $r_{\rm Auger}= r_{\rm min}$, the 3-$\mu$s delay time corresponds to a maximal e-h plasma density of about $10^{18}\,{\rm cm^{-3}}$ (black dashed line in Fig.~\ref{FIG:simulation_refl_pulse} (b)), which is right above the critical carrier density $N_{\rm Mott}$ for exciton-plasma Mott transition in Si. For longer delay times, it is thus reasonable to assume a higher Auger coefficient to account for stronger Coulomb interaction because of weaker free-carrier screening for lower e-h plasma densities~\cite{hangleiter1990enhancement}. Similar to the empirical formulae proposed in earlier works~\cite{altermatt1997assessment, jonsson1997ambipolar}, we introduce the following function imitating the Fermi function to quantify the transition of the Auger coefficient from a higher value $r_{\rm max}$ to the lower value $r_{\rm min}$:
\begin{align}
r_{\rm Auger}
=
r_{\rm min}
+
\frac
{r_{\rm max}-r_{\rm min}}
{e^{\alpha(\frac{N}{N_0}-1)}+1},
\label{EQ:r_auger}
\end{align}
where the parameters $N_0$ and $\alpha$ respectively characterize the critical carrier density and abruptness of the transition. The three parameters, $r_{\rm max}$, $N_0$, and $\alpha$, are adjusted one by one iteratively to minimize the deviation between the simulated and measured reflectance values for the pulse experiment. Excellent agreement between the theory and experiment is shown in Fig.~\ref{FIG:simulation_refl_pulse} (a), where the optimal set of parameters, $r_{\rm max}=1.65\times10^{-39}\,{\rm cm^6/ns}$, $N_0=6.31\times10^{17}\,{\rm cm^{-3}}$, and $\alpha=6$, is used to describe the $N$-dependent Auger coefficient (red solid line in Fig.~\ref{FIG:auger_coefficient}) for the simulation (red solid line in Fig.~\ref{FIG:simulation_refl_pulse} (a)). We see from the fitting results that the transitions from a weak to a strong carrier-screening regime for the Auger recombination and the e-h scattering both happen around the critical carrier density for exciton-plasma Mott transition, as one should expect.

Development of an empirical formula remains the most important way for characterizing the Auger recombination in Si because of the lack of a simple quantitative understanding of the phenomenon~\cite{altermatt1997assessment,jonsson1997ambipolar,glunz1999field,schmidt2000coulomb,kerr2002general,richter2012improved,niewelt2022reassessment,black2022quantification}. To quantify the transition from a high to a low Auger coefficient, a theory that contains no adjustable parameters does exist based on the consideration of carrier-screening effects~\cite{hangleiter1990enhancement}. However, the theory failed to reproduce the sharp Mott transition in the minority-carrier lifetimes in highly n-doped Si at 300\,K~\cite{hacker1994intrinsic}. Another complexity comes from the fact that Auger recombination can occur through various mechanisms, which include direct processes involving only electrons and holes in the energy bands~\cite{huldt1971band,hill1976formalism,laks1988theory,laks1990accurate,govoni2011auger}, phonon-assisted processes~\cite{haug1978auger,lochmann1978phonon,lochmann1978phonon,bushick2023phonon}, and trap-assisted processes involving impurities~\cite{evans1963recombination}.

To set a high-injection limit, the Auger coefficients, determined four decades ago by Dziewior and Schmid from measurements of minority-carrier lifetimes in Si with high doping levels from $10^{18}$ to $10^{20}\,{\rm cm^{-3}}$~\cite{dziewior1977auger}, have been extensively used in empirical parametrization of Auger recombination~\cite{altermatt1997assessment,jonsson1997ambipolar,glunz1999field,schmidt2000coulomb,kerr2002general,richter2012improved,niewelt2022reassessment,black2022quantification}, even in the recent reassessment of intrinsic bulk recombination time in Si~\cite{niewelt2022reassessment}. The associated ambipolar Auger coefficient, which is around $3.8\times10^{-40}\,{\rm cm^6/ns}$~\cite{dziewior1977auger} (horizontal orange line in Fig.~\ref{FIG:auger_coefficient}), has also been widely used for simulating LDSSs~\cite{vogel1992high,nozokido1997modulation,picard2019laser,li2023measurement,schaub2021laser}. The value is more than one order of magnitude larger than the minimal value $r_{\rm min}=1.8\times10^{-41}\,{\rm cm^6/ns}$ (black dashed line) determined here. On the one hand, the minority-carrier lifetimes in Si with such high doping levels have not been reassessed. On the other hand, it is reasonable to expect that the lifetime of an optically injected e-h plasma in lowly doped Si should be longer than the minority-carrier lifetime in highly doped Si with a comparable carrier concentration. Compared to highly doped Si, the screening of Coulomb interaction between free electrons and holes is more pronounced in high-density e-h plasmas in lowly doped Si because of the more balanced electric charge, as has been noted in the theoretical work in Ref.~\cite{hangleiter1990enhancement}. In addition, it is obvious that trap-assisted Auger recombination is more likely to happen in Si with a higher doping level~\cite{evans1963recombination}. The minimal ambipolar Auger coefficient $r_{\rm min}$ deduced here agrees well with the values calculated in Ref.~\cite{huldt1971band} ($2\times10^{-41}\,{\rm cm^6/ns}$) and Ref.~\cite{hill1976formalism} ($1.2\times10^{-41}\,{\rm cm^6/ns}$), in which direct Auger processes involving only free electrons and holes are considered based on the low-energy band structure of Si. If a constant ambipolar Auger coefficient $r_{\rm Auger}=3.8\times10^{-40}\,{\rm cm^6/ns}$ is used in our reflectance simulation, the reflectance values will be underestimated (by around 0.1 near the rising edge) with an optimal value of $\mu_{\rm eh,\infty}$ being $189\,\rm cm^2/(V\cdot s)$ (Fig.~\ref{FIG:simulation_refl_pulse} (a)).

For relatively low e-h plasma densities, the Auger recombination process in Si with carrier-injection levels ranging from $10^{14}$ to $10^{17}$ ${\rm cm}^{-3}$ has been reassessed recently through photoconductance-decay and modulated luminescence experiments, in which a large number of Si wafers processed by state-of-the-art surface passivation were investigated~\cite{niewelt2022reassessment}. The ambipolar Auger coefficient is parametrized by using the empirical formula proposed by Jonsson et. al.~\cite{jonsson1997ambipolar} (black solid line in Fig.~\ref{FIG:auger_coefficient}), incorporating the lifetime data from Dziewior and Schmid~\cite{dziewior1977auger} with a slight modification according to a more updated mobility model~\cite{black2022quantification}. The relative deviations of the extracted Auger recombination rates in Ref.~\cite{niewelt2022reassessment} from the empirical formula are mostly within 20\% (shade area). The Auger coefficient extracted here (red solid line) lies in this 20\% deviation window for the carrier-injection levels in the reassessment.

Our empirical formula of ambipolar Auger coefficient is also consistent with the data from the literature~\cite{pang1991record,pang1993new,sinton1987recombination,misiakos1990carrier,yablonovitch1986auger,forget1992auger,svantesson1979determination} (scattered markers in Fig.~\ref{FIG:auger_coefficient}). For the relatively high Auger recombination rate from Ref.~\cite{yablonovitch1986auger} (hexagrams), it has been argued that the Auger recombination coefficient should be lowered if trap-assisted Auger processes were included~\cite{landsberg1987trap}. For the relatively low Auger recombination rates from Ref.~\cite{pang1991record} (down-pointing triangles) and Ref.~\cite{pang1993new} (asterisks), using an ambipolar Auger coefficient of $1.7\times10^{-39}\,{\rm cm^6/ns}$ can also result in a possible interpretation of the lifetime data therein~\cite{pang1991record}.

In conclusion, we have derived a two-fluid mobility model incorporating fundamental aspects of classical e-h scattering such as momentum conservation for simulating LDSSs based on two coupled hydrodynamic equations. The two-fluid mobility model predicts a different frequency dependence in the dielectric function and thus distinct AC responses of e-h plasmas, compared to previous works that use Matthiessen's rule. Using a formula that describes classical mutual diffusion of electron and hole gases, we determine the parameters in the two-fluid mobility model for e-h plasma densities up to $10^{18}\,\rm cm^{-3}$ by showing the consistency with existing DC-field mobility data. We demonstrate a prototypical application of the two-fluid mobility model for simulating the reflectance at around 320\,GHz in a Si switch driven by a 5-ns green-laser pulse, which injects e-h plasmas with a density up to the order of $10^{20}\,\rm cm^{-3}$. We find that carrier screening effects in both the e-h scattering and Auger recombination become important for carrier densities above the critical carrier density for exciton-plasma Mott transition, which is about $10^{18}\,\rm cm^{-3}$. Two empirical formulae are introduced to characterize respectively the e-h scattering in the two-fluid mobility model and the Auger recombination for e-h plasmas with densities comparable to or above $10^{18}\,\rm cm^{-3}$. The experimentally measured reflectance up to 1\,$\mu$s after the pump pulse is well reproduced by using only two fitting parameters (five fitting parameters for reproducing the reflectance up to 60\,$\mu$s after the pump pulse). Each of the fitting parameters is unambiguously determined by examining a specific feature in the temporal profile of reflectance. By using a high-purity Si wafer with well-passivated surfaces, our results also suggest a way to characterize the intrinsic momentum-relaxation mechanism, e-h scattering, and the intrinsic e-h recombination mechanism in indirect-gap semiconductors, Auger recombination. More generally, we expect that the distinct AC responses predicted by the two-fluid mobility model will stimulate further studies of e-h plasmas in the high carrier-density regime and lead to more refined design of semiconductor devices.

\section*{Acknowledgments}

Q. W., A. S., B. D. P., N. I. A., and M. S. S. were supported by NSF-DMR 2117994 and 2333941.  J. D. M. and T. N. were funded by the Leverhulme Trust (RPG-2020-377). A. Y. was in receipt of an ISIS Facility Development Studentship from the Science and Technology Facilities Council.

Q. W. developed the theory and did all calculations. A. S. designed and conducted the reflectance measurements with support from B. D. P., N. I. A., and M. S. S.. A. Y. and S. L. P. prepared the silicon wafer and did the preliminary characterization under the supervision of J. D. M. and the administration of T. N.. Q. W. wrote the original manuscript and finalized the manuscript with support from A. S., B. D. P., S. L. P., J. D. M., T. N., and M. S. S..

The data that support the findings of this article are openly available~\cite{qile2025effective}.
\appendix

\section{Derivation of the ambipolar diffusion coefficient}\label{APP:diffusion_coef}

We start the derivation with the assumption of local charge neutrality, which means the electrons and holes have the same particle density $N$ and flux $J$ along the direction normal to the Si-wafer surfaces. Considering generation and recombination processes of electron-hole pairs, we write down the following continuity equation:
\begin{align}
\frac{\partial N}{\partial t}
=
\frac{\partial J}{\partial z}
+G(z,t)-R(N),
\end{align}
where $G(z,t)$ and $R(N)$ are the generation and recombination rates of the electron-hole pairs, respectively. Since the electrons generally diffuse faster than the holes, an electric field $E_{\rm int}$ forms between the electrons and holes such that they can have the same particle flux. Therefore, the flux $J$ contains a drift current due to the internal electric field $E_{\rm int}$ as well as a diffusion current, i.e.,
\begin{align}
J
&=-N\mu_{\rm e}(0) E_{\rm int}-D_{\rm e}\frac{\partial N}{\partial z}\notag\\
&=N\mu_{\rm h}(0) E_{\rm int}-D_{\rm h}\frac{\partial N}{\partial z},
\label{EQ:J_twoparts}
\end{align}
where $D_{\rm e}$ ($D_{\rm h}$) and $\mu_{\rm e}(0)$ ($\mu_{\rm h}(0)$) are the diffusion coefficient and DC-field mobility of the electrons (holes), respectively. The internal electric field $E_{\rm int}$ can be solved from Eq.~(\ref{EQ:J_twoparts}) as
\begin{align}
E_{\rm int}
=
\frac{D_{\rm h}-D_{\rm e}}{N [\mu_{\rm e}(0)+\mu_{\rm h}(0)]}
\frac{\partial N}{\partial z}.
\label{EQ:E_int}
\end{align}
Putting Eq.~(\ref{EQ:E_int}) back into Eq.~(\ref{EQ:J_twoparts}) leads to the expression of ambipolar diffusion current, 
\begin{align}
J=D(N)\frac{\partial N}{\partial z},
\end{align}
where $D(N)={[D_{\rm e}\mu_{\rm h}(0)+D_{\rm h}\mu_{\rm e}(0)]}/{[\mu_{\rm h}(0)+\mu_{\rm e}(0)]}$ is the ambipolar diffusion coefficient. We further assume that the electron and hole gases are both nondegenerate and are in thermal equilibrium with the lattice at temperature $T$ such that the Einstein relations $D_{\rm e}=\mu_{\rm e}(0)k_{\rm B}T/e$ and $D_{\rm h}=\mu_{\rm h}(0)k_{\rm B}T/e$ are valid~\cite{linnros1994carrier}. Here, $e$ is the elementary charge, and $k_{\rm B}$ is the Boltzmann constant. Thus, we write the ambipolar diffusion coefficient in the form:
\begin{align}
D(N)
=
\frac{2k_BT}{e}\frac{\mu_{\rm e}(0)\mu_{\rm h}(0)}{\mu_{\rm h}(0)+\mu_{\rm e}(0)}.
\label{EQ:ambi_diff}
\end{align}
As shown in Eq.~(\ref{EQ:ambi_diff}), the ambipolar diffusion coefficient is determined by both the electron and hole mobilities. Therefore, if the electron and hole mobilities are treated as independent parameters, it is not possible to separately extract the electron and hole mobilities from ambipolar diffusivity data. In the two-fluid mobility model for undoped semiconductors, the electron and hole mobilities are parametrized with the same set of parameters, which includes $\mu_{\rm eh}$ associated with e-h scattering and $\mu_{\rm e(h),0}$ associated with other scattering mechanisms for the electrons (holes). Using Eqs.~(\ref{EQ:Mobility_electron}) and~(\ref{EQ:Mobility_hole}), we can rewrite Eq.~(\ref{EQ:ambi_diff}) as
\begin{align}
D(N)
=
\frac{({2k_BT}/{e})\mu_{\rm e,0}\mu_{\rm h,0}}{(\mu_{\rm h,0}+\mu_{\rm e,0})[1+(\mu_{\rm h,0}+\mu_{\rm e,0})\mu^{-1}_{\rm eh}]}.
\label{EQ:ambi_diff_ex}
\end{align}
If the parameters $\mu_{\rm e,0}$ and $\mu_{\rm h,0}$ are taken as known constants, for example, from mobility measurements on single crystals~\cite{ludwig1956drift}, the parameter $\mu_{\rm eh}$ and thus the electron and hole mobilities can be determined from ambipolar diffusivity data by using Eq.~(\ref{EQ:ambi_diff_ex}). 

\section{Lattice contribution to the dielectric function}\label{APP:lattice_dielectric}

\begin{figure}
	\includegraphics[width=0.47\textwidth]{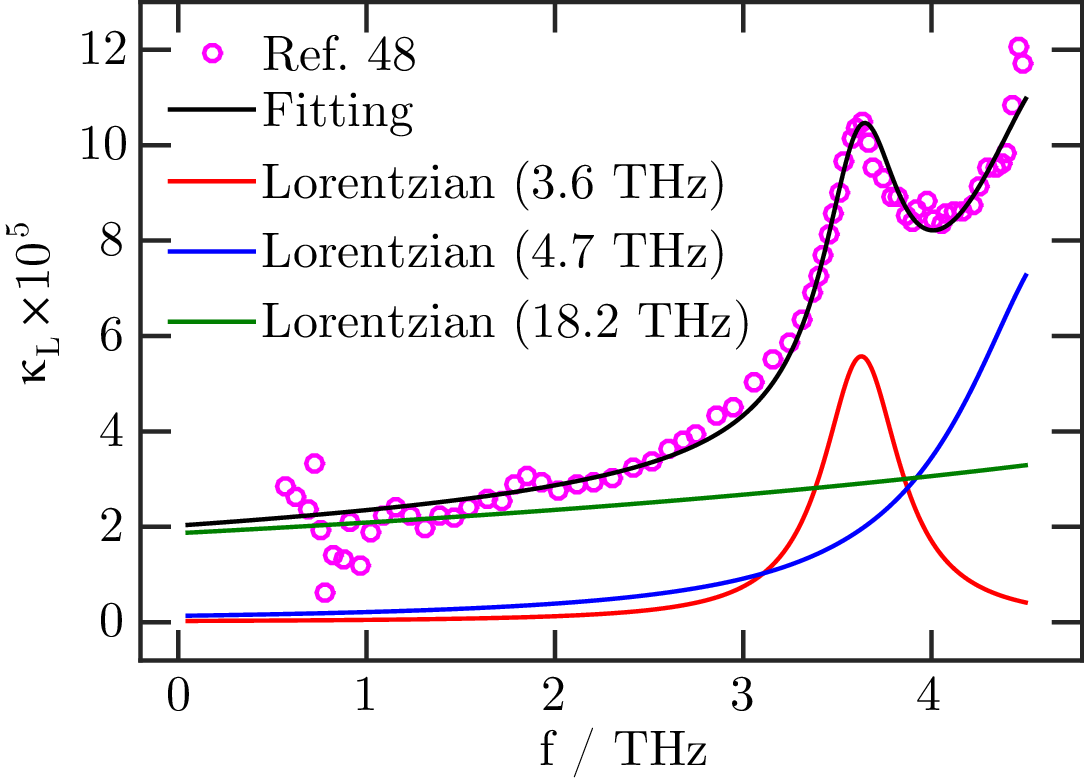}
	\caption{Extrapolating the extinction coefficient. The extinction coefficient calculated from the absorption coefficient in Ref.~\cite{dai2004terahertz} is shown as magenta circles. A sum (black curve) of three Lorentzian functions peaked at around 3.6 THz (red), 4.7 THz (blue), and 18.2 THz (dark green), respectively, is fit to the data.}
	\label{FIG:phonon_extinction}
\end{figure}

The contribution to the dielectric function from lattice vibrations for the sub-THz waves, $\epsilon_{\rm L}$, is estimated from the data in a frequency range from 0.5 to 4.5\,THz in Ref.~\cite{dai2004terahertz}, where the refractive index $n_{\rm L}$ was shown to be approximately constant at around 3.4175. We simply take $n_{\rm L}=3.4175$ for our 314.4-GHz waves. The extinction coefficient $\kappa_{\rm L}$ calculated from the absorption coefficient in Ref.~\cite{dai2004terahertz} is shown in Fig.~\ref{FIG:phonon_extinction}. To extrapolate the data to cover the lower-frequency domain, we use three Lorentzian functions with peak positions respectively at 121 ${\rm cm}^{-1}$ ($f_1\approx3.6$\,THz), 157 ${\rm cm}^{-1}$ ($f_2\approx4.7$\,THz), and 608 ${\rm cm}^{-1}$ ($f_3\approx18.2$\,THz), corresponding to three two-phonon processes~\cite{ikezawa1981far,abdullah1984observation}. As shown in Fig.~\ref{FIG:phonon_extinction}, a sum of three Lorentzian functions, $\sum_{j=1}^3{Y_j}/{[(f-f_j)^2+\Gamma_j]}$ is well fitted to the data. Here, the fitting parameters are: $Y_1=3.4\times10^{-6}$,
$Y_2=3.0\times10^{-5}$, $Y_3=6.2\times10^{-3}$, $\Gamma_1=0.061$, $\Gamma_2=0.37$, and $\Gamma_3=0$. We take $\kappa_{\rm L}=2\times10^{-5}$ based on the fitting result.

\section{Reflectance measurements}\label{APP:reflectance_exp}

\begin{figure}
	\includegraphics[width=0.47\textwidth]{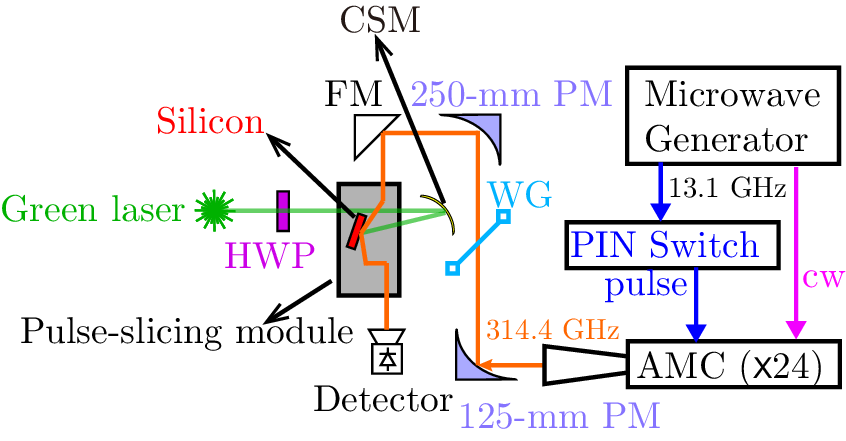}
	\caption{Schematic diagram of the experimental setup for the relfectance measurements. AMC: amplifier/multiplier chain. cw: continuous wave. WG: wire-grid polarizer. PM: parabolic mirror. FM: flat mirror. HWP: half-wave plate. CSM: concave spherical mirror.}
	\label{FIG:measurement_setup}
\end{figure}
The experimental setup for the reflectance measurements is schematically shown in Fig.~\ref{FIG:measurement_setup}. Continuous electromagnetic waves with a frequency of 13.1\,GHz were generated by a microwave generator (SynthHD Mini, Windfreak Technologies, LLC USA) and then propagated through an amplifier/multiplier chain (AMC) (Custom made AMC, Virginia Diodes, Inc., USA), after which the frequency of the microwaves was multiplied by 24 to be a sub-THz frequency of 314.4\,GHz. The frequency of the input microwaves was chosen to achieve a high output power up to about 10\,mW after the frequency multiplication. The sub-THz waves were launched through a taper of 8\,mm in diameter into the air, focused by two parabolic mirrors with focal lengths of 125\,mm and 250\,mm, respectively, and then directed by a flat mirror into a pulse-slicing module~\cite{price2024compact}, which passed the sub-THz waves onto the silicon (Si) wafer through a parabolic mirror and guided the reflected waves into a quasi-optical detector (3DL 12C LS2500 A2,ACST GmbH, DE) through several parabolic mirrors (see Ref.~\cite{price2024compact} for a detailed description of the module). A wire-grid polarizer was placed between the 125-mm and 250-mm focusing parabolic mirrors to make the electric fields of the sub-THz waves lie in the plane of incidence with respect to the top surface of the silicon (Si) wafer (p-polarization). The green-laser beams (DPS-532-BS-D-50mJ, Changchun New Industries Optoelectronics Tech. Co., Ltd., CN) were sent from outside of the pulse-slicing module and expanded by a concave spherical mirror to an elliptical shape with the major and minor axes being about 50 and 35\,mm, respectively, to fully cover the sub-THz beam spot on the Si wafer. The sub-THz beam spot on the Si-wafer surface was also in an elliptical shape, whose major and minor axes were estimated to be about 46 and 8\,mm, respectively. A half-wave plate was used to tune the electric fields of the green-laser pulses to p-polarization to maximize the carrier-generation rate. In the reflectance measurements with sub-THz pulses, a PIN diode-based switch was placed between the microwave generator and AMC to generate 24-ns sub-THz pulses.

\section{Normalization of reflectance}\label{APP:r_normalization}

\begin{figure}
	\includegraphics[width=0.47\textwidth]{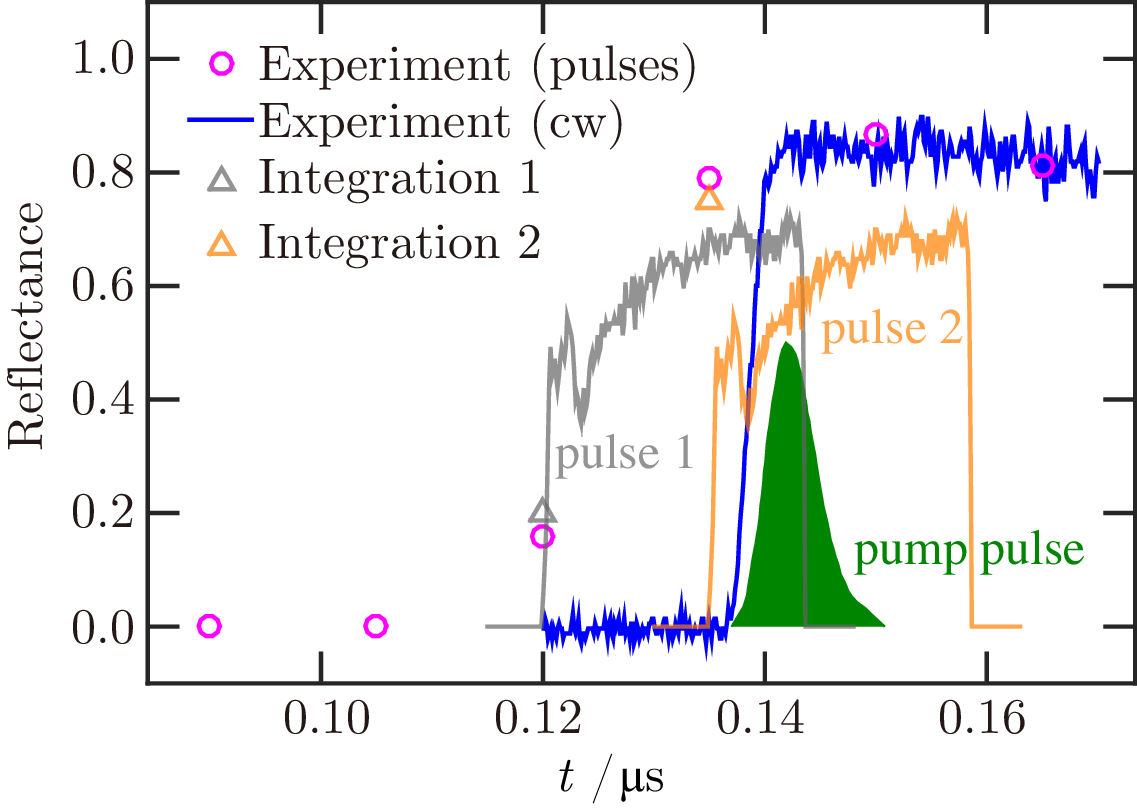}
	\caption{Normalization of reflectance. The magenta circles represent six averaged reflectance values measured by using 25-ns sub-THz pulses. The delay times of the sub-THz pulses with respect to the green-laser pump pulse are spaced by 15\,ns. The grey and orange curves show the temporal profiles of two sub-THz pulses that correspond to the magenta circles at around 0.12\,$\mu$s and 0.135\,$\mu$s, respectively. The blue curve shows the reflectance measured with continuous sub-THz waves after normalization. The grey (orange) triangle represents the value of the overlap integral between the temporal profiles of the sub-THz pulse 1 (2) and the reflectance measured with continuous sub-THz waves. The dark-green shaded area shows the temporal profile of the green-laser pulse.}
	\label{FIG:r_normalization}
\end{figure}

For the reflectance measurements with sub-THz pulses, both the temporal profiles of the input and reflected pulses were measured, and each of the reflectance value shown in this paper was taken as the total energy in the reflected pulse normalized with respect to the total energy in the input pulse. To measure the temporal profile of an input sub-THz pulse, the Si wafer was removed from the optical path of the sub-THz wave in the pulse-slicing module, and a flat mirror was inserted to direct the sub-THz pulse into the quasi-optical detector. Figure~\ref{FIG:r_normalization} shows six reflectance values (magenta circles) measured by using sub THz-pulses. For example, the input pulse rising at around 0.12\,$\mu$s (pulse 1, grey curve) corresponds to the reflectance value of around 0.2. 

For the reflectance measurements with continuous sub-THz waves, because the quasi-optical detector is only sensitive to the changes of input power, one can only measure a relative reflectance $R^{\rm cw}_{\rm raw}(t)$, which differs from the actual reflectance by a constant factor $A_{\rm cw}$. Sending a continuous-wave (cw) sub-THz beam directly to the detector will result in a null signal. To obtain the actual reflectance $A_{\rm cw}R^{\rm cw}_{\rm raw}(t)$ from the cw experiment, we make use of its connection with the reflectance values from the pulse experiment. Each reflectance value from the pulse experiment is ideally the same as the overlap integral between the temporal profiles of the corresponding sub-THz pulse and the actual reflectance, i.e.,
\begin{align}
R^{\rm ideal}_{\rm pulse\,j}
=
&\frac{\int_{-\infty}^{+\infty} dt h_{\rm pulse\,j}(t)A_{\rm cw}R^{\rm cw}_{\rm raw}(t)}
{\int_{-\infty}^{+\infty} dt h_{\rm pulse\,j}(t)}
\notag\\
&\equiv A_{\rm cw}R^{\rm raw}_{\rm pulse\,j},
\label{EQ:overlap_int}
\end{align}
where $h_{\rm pulse\,j}(t)$ is the temporal profile of the j-th sub-THz pulse. The reflectance values $R_{\rm pulse\,j}$ (j=1,2) from the pulse experiment with pulses 1 and 2 (grey and orange curves in Fig.~\ref{FIG:r_normalization}) are used to determine the normalization constant $A_{\rm cw}$. We minimize the total squared deviation $\sum_{j=1,2}|A_{\rm cw}-R_{\rm pulse\,j}/R^{\rm raw}_{\rm pulse\,j}|^2$ and calculate $A_{\rm cw}$ as
\begin{align}
A_{\rm cw}=\frac{1}{2}\sum_{j=1,2}\frac{R_{\rm pulse\,j}}{R^{\rm raw}_{\rm pulse\,j}},
\end{align}
where $R^{\rm raw}_{\rm pulse\,j}$ is calculated from Eq.~(\ref{EQ:overlap_int}). The values of $R^{\rm ideal}_{\rm pulse\,j}$ for pulses 1 and 2 are shown as grey and orange triangles, respectively, in Fig.~\ref{FIG:r_normalization}.

\section{Light propagation in stratified media}\label{APP:light_propagation}
\begin{figure}
	\includegraphics[width=0.47\textwidth]{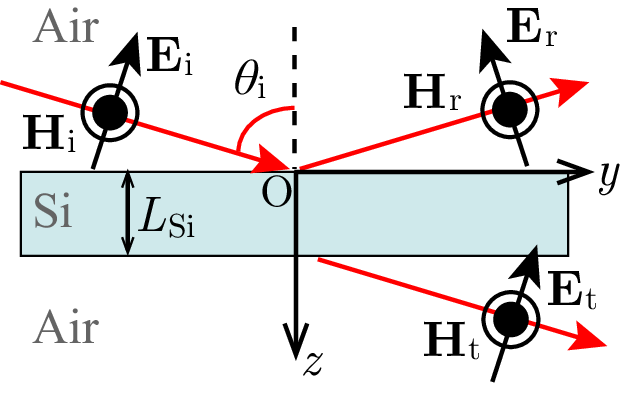}
	\caption{Coordinate system for solving the wave equations. A plane wave is incident at an angle $\theta_{\rm i}$ onto a silicon (Si) wafer of thickness $L_{\rm Si}$. The magnetic field $\bf H$ is parallel to the $x$-axis which points into the paper plane, while the electric field $\bf E$ lies in the $yz$-plane.}
	\label{FIG:transfer_matrix}
\end{figure}
Following the general treatment of light propagation in stratified media in Ref.~\cite{born2013principles}, we derive here the transfer matrix for calculating the electric field of the green light in the Si wafer and the reflectance for the 314.4-GHz waves. The incident light waves in this paper are all p-polarized with respect to the top surface of the Si wafer, i.e., their electric fields lie in the plane of incidence. Consider a monochromatic plane wave propagating in an isotropic slab of a linear material characterized by a complex permittivity $\varepsilon$ and a permeability $\mu\approx1$. In the absence of free charges, the wave equations for the electric field $\bf E$ and the magnetic field $\bf H$ can be written as
\begin{align}
\nabla\times {\bf E}=i\mu_0\omega {\bf H},
\\
\nabla\times {\bf H}
=
-i\varepsilon_0\varepsilon\omega{\bf E},
\end{align}
where $\omega$ is the angular frequency of the plane wave, $\varepsilon_0$ is the vacuum permittivity, and $\mu_0$ is the vacuum permeability. In the coordinate system shown in Fig.~\ref{FIG:transfer_matrix}, where the magnetic field is along the $x$-axis ($H_y=H_z=0$) and the electric field lies in the $yz$-plane ($E_x=0$), the above wave equations explicitly read
\begin{align}
\partial_y E_z-\partial_z E_y
=
i\mu_0\omega H_x,
\\
\partial_x E_z=\partial_x E_y=0,
\\
\partial_z H_x
=
-i\varepsilon_0\varepsilon\omega E_y,
\label{EQ:Hx_Ey}
\\
-\partial_y H_x
=
-i\varepsilon_0\varepsilon\omega E_z.
\label{EQ:Hx_Ez}
\end{align}
The solution compatible with a plane-wave input from the air has the following form:
\begin{align}
H_x=U(z)\exp[i(\beta k_0 y-\omega t)],\\
E_y=-V(z)\exp[i(\beta k_0 y-\omega t)],\\
E_z=-W(z)\exp[i(\beta k_0 y-\omega t)],
\end{align}
where $\beta=n_{\rm Air}\sin\theta_{\rm i}$ is defined by the refractive index of the air $n_{\rm Air}=1.0003$~\cite{ciddor1996refractive} and the angle of incidence $\theta_{\rm i}$, $k_0$ is the wave vector of the plane wave in vacuum, and the auxiliary functions, $U(z)$, $V(z)$, and $W(z)$, satisfy
\begin{align}
U''(z)=-k_0^2(\varepsilon-\beta^2) U(z),
\label{EQ:U_z}
\\
V(z)=\frac{1}{i\omega\varepsilon_0\varepsilon}U'(z),
\label{EQ:Uz_Vz}
\\
W(z)=-\sqrt{\frac{\mu_0}{\varepsilon_0}}\frac{\beta}{\varepsilon} U(z).
\label{EQ:Uz_Wz}
\end{align}
From Eqs.~(\ref{EQ:U_z}) and~(\ref{EQ:Uz_Vz}), the general solutions for $U(z)$ and $V(z)$ can be easily obtained as
\begin{align}
\begin{pmatrix}
U(z)\\
V(z)
\end{pmatrix}
=
\mathbb{M}
\begin{pmatrix}
U(0)\\
V(0)
\end{pmatrix}
=
\begin{pmatrix}
\mathbb{M}_{11}  &\mathbb{M}_{12}\\
\mathbb{M}_{21} & \mathbb{M}_{22}
\end{pmatrix}
\begin{pmatrix}
U(0)\\
V(0)
\end{pmatrix},
\label{EQ:transfer_mat}
\end{align}
where $\mathbb{M}$ is the transfer matrix defined by
\begin{align}
&\mathbb{M}_{11}
=
\mathbb{M}_{22}
=
\cos(k_0n_{\varepsilon}z) ,
\label{EQ:transfer_detail1}
\\
&
\mathbb{M}_{12}
=
i\frac{\varepsilon}{n_{\varepsilon}}
\sqrt{\frac{\varepsilon_0}{\mu_0}}
\sin(k_0n_{\varepsilon}z),
\label{EQ:transfer_detail2}
\\
&
\mathbb{M}_{21}
=
i\frac{n_{\varepsilon}}{\varepsilon}
\sqrt{\frac{\mu_0}{\varepsilon_0}}
\sin(k_0n_{\varepsilon} z),
\label{EQ:transfer_detail3}
\end{align}
with $n_{\varepsilon}\equiv\sqrt{\varepsilon-\beta^2}$. In this paper, a square root of a complex number is defined to have a nonnegative real part. If the amplitudes of the electromagnetic fields are known on one surface of the slab, the electromagnetic fields at any point in the material can be calculated through Eqs.~(\ref{EQ:transfer_mat}) and~(\ref{EQ:Uz_Wz}).

\section{Absorptive dissipation of the green light in the Si wafer}\label{APP:green_light}
From the transfer matrix formalism derived in the last section, one can calculate the absorptive dissipation of the green light in the Si wafer, if the light field on the top surface of the Si wafer ($z=0$) is known. In this section, the dielectric constant $\varepsilon$ is understood as the dielectric constant of silicon for the green light at 532.263\,nm, and $\theta_{\rm i}=45^{\circ}$ is the incident angle. In the air above the Si wafer in Fig.~\ref{FIG:transfer_matrix}, the magnetic fields of the incident and reflected waves take the simple forms,
\begin{align}
{\bf H}_{\rm i}
=
&\hat{x}
H_{\rm i}\notag\\
&\times
\exp\{i[n_{\rm Air}(\sin\theta_{\rm i}y+\cos\theta_{\rm i}z) k_{\rm gr} -\omega_{\rm gr} t]\},
\label{EQ:incident_H}\\
{\bf H}_{\rm r}
=
&\hat{x}
H_{\rm r}\notag\\
&\times
\exp\{i[n_{\rm Air}(\sin\theta_{\rm i}y+\cos\theta_{\rm i}z) k_{\rm gr} -\omega_{\rm gr} t]\},
\end{align}
which can be seen by solving Eq.~(\ref{EQ:U_z}) with $\varepsilon$ taken to be $n_{\rm Air}^2$. Here, $k_{\rm gr}$ and $\omega_{\rm gr}$ denote the free-space wavevector and angular frequency of the green laser. The corresponding electric fields can be calculated directly from Eqs.~(\ref{EQ:Hx_Ey}) and~(\ref{EQ:Hx_Ez}) as
\begin{align}
{\bf E}_{\rm i}
=
&\frac{H_{\rm i}(\hat{z}\sin\theta_{\rm i}-\hat{y}
\cos\theta_{\rm i})}{n_{\rm Air}\sqrt{\varepsilon_0/\mu_0}}\notag\\
&\times
\exp\{i[n_{\rm Air}(\sin\theta_{\rm i}y+\cos\theta_{\rm i}z) k_{\rm gr} -\omega_{\rm gr} t]\},
\label{EQ:incident_E}\\
{\bf E}_{\rm r}
=
&\frac{H_{\rm r}(\hat{y}
\cos\theta_{\rm i}+\hat{z}\sin\theta_{\rm i})}{n_{\rm Air}\sqrt{\varepsilon_0/\mu_0}}\notag\\
&\times
\exp\{i[n_{\rm Air}(\sin\theta_{\rm i}y-\cos\theta_{\rm i}z) k_{\rm gr} -\omega_{\rm gr} t]\}.
\end{align}
The total light field in the air above the Si wafer is a superposition of the incident field and the reflected field. On the top surface of the Si wafer ($z=0$), The auxiliary functions $U(z)$ and $V(z)$ have the values:
\begin{align}
&U(0)
=
H_{\rm i}+H_{\rm r},
\label{EQ:gr_u0}
\\
&V(0)
=
\frac{\cos\theta_{\rm i}}{n_{\rm Air}\sqrt{\varepsilon_0/\mu_0}}
(H_{\rm i}-H_{\rm r}).
\label{EQ:gr_v0}
\end{align}
The electromagnetic fields in the Si wafer can then be calculated from Eqs.~(\ref{EQ:transfer_mat}) and~(\ref{EQ:Uz_Wz}), which give
\begin{align}
U(z)
=
&(H_{\rm i}+H_{\rm r})\cos(k_{\rm gr}n_{\varepsilon}z)\notag\\
&+
(H_{\rm i}-H_{\rm r})\frac{i\varepsilon\cos\theta_{\rm i}}{n_{\rm Air}n_{\varepsilon}}
\sin(k_{\rm gr}n_{\varepsilon}z),\\
V(z)
=
&i\frac{n_{\varepsilon}}{\varepsilon}\sqrt{\frac{\mu_0}{\varepsilon_0}}
[
(H_{\rm i}+H_{\rm r})
\sin(k_{\rm gr}n_{\varepsilon} z)\notag\\
&-
(H_{\rm i}-H_{\rm r})\frac{i\varepsilon\cos\theta_{\rm i}}{n_{\rm Air}n_{\varepsilon}}
\cos(k_{\rm gr}n_{\varepsilon} z)
],
\label{EQ:E_Vz}
\\
W(z)
=
&
-\frac{n_{\rm Air}\sin\theta_{\rm i}}{\varepsilon}\sqrt{\frac{\mu_0}{\varepsilon_0}}
[(H_{\rm i}+H_{\rm r})\cos(k_{\rm gr}n_{\varepsilon}z)\notag\\
&+
(H_{\rm i}-H_{\rm r})\frac{i\varepsilon\cos\theta_{\rm i}}{n_{\rm Air}n_{\varepsilon}}
\sin(k_{\rm gr}n_{\varepsilon}z)].
\label{EQ:E_Wz}
\end{align}
The two parameters, $H_{\rm i}$ and $H_{\rm r}$, are connected through the reflection coefficient $r_{\rm gr}=H_{\rm r}/H_{\rm i}$, which can be calculated by considering the electromagnetic fields in the air below the Si wafer in Fig.~\ref{FIG:transfer_matrix}, i.e., the transmitted fields. Similar to the incident field, the transmitted magnetic field and electric field take the following forms:
\begin{align}
{\bf H}_{\rm t}
=
&\hat{x}
H_{\rm t}\notag\\
&\times
\exp\{i[n_{\rm Air}(\sin\theta_{\rm i}y+\cos\theta_{\rm i}z) k_{\rm gr} -\omega_{\rm gr} t]\},
\label{EQ:gr_u}
\\
{\bf E}_{\rm t}
=
&\frac{H_{\rm t}(\hat{z}\sin\theta_{\rm i}-\hat{y}
\cos\theta_{\rm i})}{n_{\rm Air}\sqrt{\varepsilon_0/\mu_0}}\notag\\
&\times
\exp\{i[n_{\rm Air}(\sin\theta_{\rm i}y+\cos\theta_{\rm i}z) k_{\rm gr} -\omega_{\rm gr} t]\}.
\label{EQ:gr_vw}
\end{align}
Continuity of the electromagnetic fields across the bottom surface ($z=L_{\rm Si}$) requires that
\begin{align}
&
H_{\rm t}
\exp({i n_{\rm Air}\cos\theta_{\rm i} k_{\rm gr}L_{\rm Si}})
=
U(L_{\rm Si})\notag\\
=&
(H_{\rm i}+H_{\rm r})\cos(k_{\rm gr}n_{\varepsilon}L_{\rm Si})\notag\\
&+
(H_{\rm i}-H_{\rm r})\frac{i\varepsilon\cos\theta_{\rm i}}{n_{\rm Air}n_{\varepsilon}}
\sin(k_{\rm gr}n_{\varepsilon}L_{\rm Si}),\\
&
H_{\rm t}
\exp({i n_{\rm Air}\cos\theta_{\rm i} k_{\rm gr}L_{\rm Si}})
\frac{
\cos\theta_{\rm i}
}
{n_{\rm Air}\sqrt{\varepsilon_0/\mu_0}}
=
V(L_{\rm Si})\notag\\
=&
\sqrt{\frac{\mu_0}{\varepsilon_0}}
[
(H_{\rm i}+H_{\rm r})i\frac{n_{\varepsilon}}{\varepsilon}
\sin(k_{\rm gr}n_{\varepsilon} L_{\rm Si})\notag\\
&+
(H_{\rm i}-H_{\rm r})\frac{\cos\theta_{\rm i}}{n_{\rm Air}}
\cos(k_{\rm gr}n_{\varepsilon} L_{\rm Si})
],
\end{align}
from which one can solve the reflection coefficient $r_{\rm gr}$ as
\begin{align}
r_{\rm gr}
=
\frac
{
i
(
\frac{n_{\rm Air}n_{\varepsilon}}{\varepsilon\cos\theta_{\rm i}}
-
\frac{\cos\theta_{\rm i}\varepsilon}{n_{\rm Air}n_{\varepsilon}}
)
\tan(k_{\rm gr}n_{\varepsilon} L_{\rm Si})
}
{
2
-
i
(
\frac{n_{\rm Air}n_{\varepsilon}}{\varepsilon\cos\theta_{\rm i}}
+
\frac{\cos\theta_{\rm i}\varepsilon}{n_{\rm Air}n_{\varepsilon}}
)
\tan(k_{\rm gr}n_{\varepsilon} L_{\rm Si})
}.
\label{EQ:reflection_coeff}
\end{align}
The parameter $H_{\rm i}$ is determined by the intensity of the incident light field. From Eqs.~(\ref{EQ:incident_E}) and~(\ref{EQ:incident_H}), the intensity of the incident light field can be calculated as an average of the Poynting vector over time, i.e.,
\begin{align}
I_{\rm in}
&
=
|\int_0^{\frac{2\pi}{\omega_{\rm gr}}}\frac{\omega_{\rm gr} dt}{2\pi}{\rm Re}({\bf E}_{\rm i})\times {\rm Re}({\bf H}_{\rm i})|
\notag\\
&
=
\frac{|H_i|^2}{2n_{\rm Air}}\sqrt{\frac{\mu_0}{\varepsilon_0}}.
\end{align}
Using Eqs.~(\ref{EQ:E_Vz}) and~(\ref{EQ:E_Wz}), we calculate the energy-density loss rate of the green-laser pulse in the absorptive dissipation in the Si wafer as
\begin{align}
\eta_{\rm abs} & (z,t)
=
\omega_{\rm gr}\varepsilon_0
{\rm Im}(\varepsilon)\int_{t-\frac{\pi}{\omega_{\rm gr}}}^{t+\frac{\pi}{\omega_{\rm gr}}}\frac{\omega_{\rm gr} dt'}{2\pi}\{{\rm Re}[{\bf E}_{\rm Si}(z,t')]\}^2\notag\\
\approx&
\frac{\omega_{\rm gr}\varepsilon_0 n_{\rm gr} \kappa_{\rm gr}}{2}(|V(z)|^2+|W(z)|^2)f^2_{\rm gr}(t),
\end{align}
where the dielectric constant $\varepsilon=(n_{\rm gr}+i \kappa_{\rm gr})^2$ is defined by the refractive index $n_{\rm gr}=4.1432$ and the extinction coefficient $\kappa_{\rm gr}=0.0324$~\cite{green2008self}, and ${\bf E}_{\rm Si}(z,t)=-f_{\rm gr}(t)[V(z)\hat{y}+W(z)\hat{z}]\exp[{i(\beta k_{\rm gr} y-\omega t)}]$ is the electric field of the green light in the Si wafer with $f^2_{\rm gr}(t)$ being the temporal profile of the green-laser pulse. In the simulation, we normalize $f^2_{\rm gr}(t)$ with respect to its maximum, so the parameter $I_{\rm in}$ corresponds to the peak intensity of the green-laser pulse. We have ignored the time for the green light to propagate through the Si wafer (about 10 ps). From Eqs.~(\ref{EQ:E_Vz}) and~(\ref{EQ:E_Wz}), the electric field of the green light attenuates in the silicon as $\exp{[- z/(2\delta_p)]}$, where $\delta_{\rm p}=[2k_{\rm gr}{\rm Im}(n_{\varepsilon})]^{-1}\approx1.3\,\mu$m is penetration depth. The thickness of the Si wafer used in this paper is $L_{\rm Si}=725\,\mu$m, which is much greater than the penetration depth $\delta_p$. Therefore, multiple reflections of the green light in the Si wafer are negligible, and the reflection coefficient $r_{\rm gr}$ of the Si wafer for the green light is close to that of an infinitely thick Si wafer. More precisely, because $k_{\rm gr}{\rm Im}(n_{\varepsilon}) L_{\rm Si}\gg1$, from Eq.~(\ref{EQ:reflection_coeff}), the reflection coefficient $r_{\rm gr}$ can be approximated as
\begin{align}
r_{\rm gr}
=
\frac
{
\varepsilon\cos\theta_{\rm i}
-
n_{\rm Air}n_{\varepsilon}
}
{
\varepsilon\cos\theta_{\rm i}
+
n_{\rm Air}n_{\varepsilon}
},
\end{align}
which corresponds to a reflectance value of $|r_{\rm gr}|^2\approx0.2466$. If all green-laser photons going into the Si wafer are absorbed and converted to e-h pairs, a green-laser pulse with a 5-ns full-width at half maximum and an intensity of $I_{\rm in}=4\,\rm MW/cm^2$ will result in a maximal e-h plasma density on the order of $(1-|r_{\rm gr}|^2)I_{\rm in}\times 5\,{\rm ns}/(\delta_p E_{\rm gr})\sim10^{20}\,{\rm cm^{-3}}$, where $E_{\rm gr}\approx 2.33\,{\rm eV}$ is the photon energy associated with the wavelength 532.263\,nm.

\section{Preliminary characterization of e-h recombination time}\label{APP:si_lifetime}
\begin{figure}
	\includegraphics[width=0.47\textwidth]{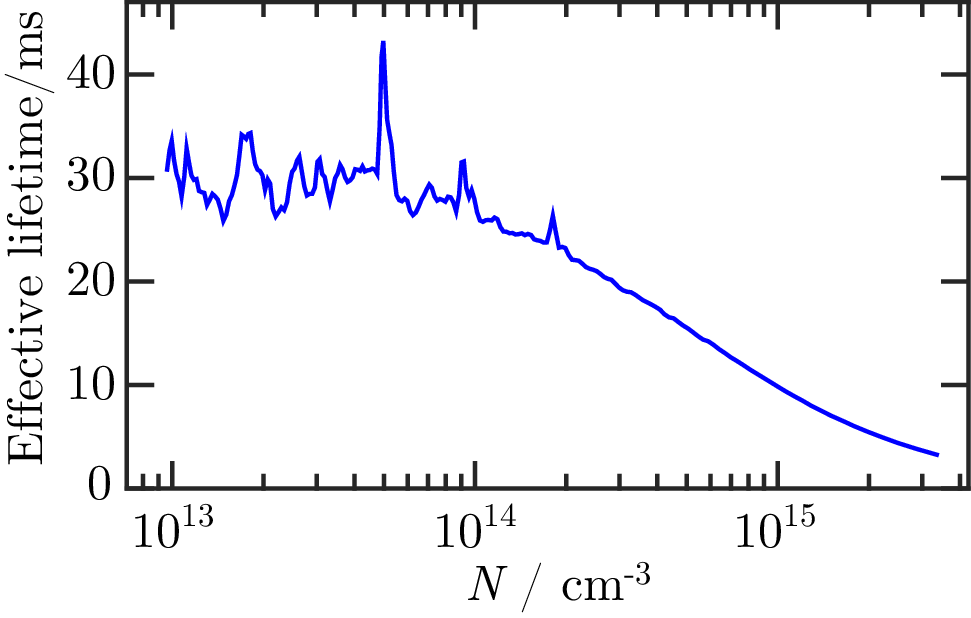}
	\caption{Effective e-h recombination times in the Si wafer at low carrier densities. The e-h recombination time is characterized as a function of the injected e-h plasma density $N$ through photoconductance-decay measurements.}
	\label{FIG:effective_lifetime}
\end{figure}

Preliminary characterization of e-h recombination time in the Si wafer was done through photoconductance-decay measurements with a Sinton WCT-120 lifetime tester~\cite{sinton1996contactless}. As shown in Fig.~\ref{FIG:effective_lifetime}, the effective e-h recombination time is not sensitive to changes in the e-h plasma density $N$ for $N<10^{14}\,{\rm cm^{-3}}$. This is consistent with a description of e-h recombination in this carrier-density range as dominated by the Shockley-Read-Hall (SRH) recombination, whose rate is proportional to the e-h plasma density $N$. From the data shown in Fig.~\ref{FIG:effective_lifetime}, we estimate the Shockley-Read-Hall (SRH) recombination time to be about $30\,{\rm ms}$, much longer than the time window of the reflectance data presented in this paper.

\section{Numerical calculation of e-h plasma density}\label{APP:numerical_plasma}

To obtain the electron-hole (e-h) plasma density in the silicon wafer, following the Crank-Nicolson method~\cite{crank1947practical}, we discretize the continuity equation,
\begin{align}
\frac{\partial N}{\partial t}
=
\frac{\partial }{\partial z}[D(N)\frac{\partial N}{\partial z}]
+G(z,t)-R(N),
\end{align}
as
\begin{align}
&
\frac{N^z_{t+\Delta t}-N^z_{t}}
{\Delta t}
\notag\\
=
&
\frac{[D(N^{z+\Delta z}_t)-D(N^{z-\Delta z}_t)][N^{z+\Delta z}_t-N^{z-\Delta z}_t]}{4\Delta z^2}
\notag\\
&
+\frac{D(N^{z}_t)}{2\Delta z^2}[N^{z+\Delta z}_{t+\Delta t}+N^{z-\Delta z}_{t+\Delta t}-2N^{z}_{t+\Delta t}\notag\\
&+N^{z+\Delta z}_{t}+N^{z-\Delta z}_{t}-2N^{z}_t]
\notag\\
&+G^z_t-R(N^{z}_t),
\label{EQ:discretization}
\end{align}
where the subscripts label the discrete times in an equidistant grid $t=0,\Delta t, 2\Delta t...$, and the superscripts label the equidistant discrete coordinates along the $z$-axis that is normal to the Si-wafer surfaces, $z=0,\Delta z,2\Delta z, ..., (q+1)\Delta z \, (q>1)$, with $(q+1)\Delta z$ being the thickness of the Si wafer. Ignoring surface recombination, we take the boundary conditions, $N^{0}_{t}=N^{\Delta z}_{t}$ and $N^{q\Delta z}_{t}=N^{(q+1)\Delta z}_{t}$.  In a matrix form, Eq.~(\ref{EQ:discretization}) can then be written as a linear equation for a $q$-dimensional vector ${\bf N}_{t+\Delta t}\equiv(N^{\Delta z}_{t+\Delta t},N^{2\Delta z}_{t+\Delta t},...,N^{q\Delta z}_{t+\Delta t})^T$,
\begin{align}
{M}_{t}{\bf N}_{t+\Delta t}={\bf u}_{t},
\label{EQ:t1_to_t2}
\end{align}
where ${\bf u}_{t}=({u}_{t,1},{u}_{t,2},...,{u}_{t,q})^T$ is a $q$-dimensional vector defined by 
\begin{align}
{u}_{t,j}
=
& \frac{\Delta t}{4\Delta z^2}[D(N^{(j+1)\Delta z}_t)-D(N^{(j-1)\Delta z}_t)]\notag\\
&\times [N^{(j+1)\Delta z}_t-N^{(j-1)\Delta z}_t]
\notag\\
&+D(N^{j\Delta_z}_t)\frac{\Delta t}{2\Delta z^2}[N^{(j+1)\Delta_z}_{t}+N^{(j-1)\Delta_z}_{t}-2N^{j\Delta_z}_t]
\notag\\
&+N^{j\Delta_z}_{t}+[G^{j\Delta_z}_t-R(N^{j\Delta_z}_t)]\Delta t,
\end{align}
and ${M}_{t}$ is a tridiagonal matrix in the form
\begin{align}
{M}_{t}
=
\begin{pmatrix}
b_{t,1}  & a_{t,1} &            &          & 0\\
a_{t,2}  & b_{t,2} & a_{t,2}    &          & \\
        & a_{t,3} & b_{t,3}     & \ddots &  \\
        &        & \ddots & \ddots & a_{t,q-1} \\
0      &        &            & a_{t,q}    & b_{t,q} \\
\end{pmatrix},
\end{align}
with $a_{t,j}=-D(N^{j\Delta z}_t){\Delta t}/({2\Delta z^2})$, $b_{t,j}=1-(2-\delta_{j,1}-\delta_{j,q})a_{t,j}$ ($j=1,2,...,q$). Both the matrix $M_t$ and the vector ${\bf u}_{t}$ are determined by the e-h plasma density $N$ at time $t$. The e-h plasma density $N$ at time $t+\Delta t$ is calculated by solving Eq.~(\ref{EQ:t1_to_t2}) through Gaussian elimination. The initial condition is set as ${\bf N}_{t=0}={\bf 0}$ with thermal excitation of e-h pairs ignored. In the calculation, we use $\Delta t=0.1\,\rm ns$ and $\Delta z=0.1\,\rm\mu m$.

\section{Reflectance of the optically excited Si wafer}\label{APP:r_subthz}

To calculate the reflectance of the optically excited Si wafer for the sub-THz waves, we need to consider a $z$-dependent dielectric function because of the attenuation of the green light in the Si wafer. As in the numerical calculation of e-h plasma density, we divide the Si wafer into $q+1$ slices perpendicular to the $z$-axis with the same thickness $\Delta z$. Within the $j$-th layer, the dielectric function is calculated as
\begin{align}
\varepsilon_j
=
\varepsilon_{\rm L}
+
i\frac{e\langle N_j[\mu_{\rm e}(N_j)+\mu_{\rm h}(N_j)]\rangle }{\varepsilon_0\omega}
,
\label{EQ:dielectric_function}
\end{align}
which includes an average over two boundary surfaces of the layer,
\begin{align}
&\langle N_j[\mu_{\rm e}(N_j)+\mu_{\rm h}(N_j)]\rangle\notag\\
=
&
\frac{1}{2}\{N^{j\Delta z}[\mu_{\rm e}(N^{j\Delta z})+\mu_{\rm h}(N^{j\Delta z})]\notag\\
&+N^{(j-1)\Delta z}[\mu_{\rm e}(N^{(j-1)\Delta z})+\mu_{\rm h}(N^{(j-1)\Delta z})\}.
\end{align}
Here, $\omega$ is the frequency of the sub-THz fields. According to Eqs.~(\ref{EQ:transfer_mat})--~(\ref{EQ:transfer_detail3}), the auxiliary functions $U(z)$ and $V(z)$ for the sub-THz fields on the top and bottom surfaces of the $j$-th layer have the following connection:
\begin{align}
\begin{pmatrix}
U(j\Delta z)\\
V(j\Delta z)
\end{pmatrix}
=&
\mathbb{M}_j
\begin{pmatrix}
U[(j-1)\Delta z]\\
V[(j-1)\Delta z]
\end{pmatrix}\notag\\
=&
\begin{pmatrix}
\mathbb{M}_{j,11}  &\mathbb{M}_{j,12}\\
\mathbb{M}_{j,21} & \mathbb{M}_{j,22}
\end{pmatrix}
\begin{pmatrix}
U[(j-1)\Delta z]\\
V[(j-1)\Delta z]
\end{pmatrix},
\label{EQ:connection_tb}
\end{align}
with the matrix elements defined as
\begin{align}
&\mathbb{M}_{j,11}
=
\mathbb{M}_{j,22}
=
\cos(k_{\rm THz}n_{\varepsilon_j}\Delta z) ,
\\
&
\mathbb{M}_{j,12}
=
i\frac{\varepsilon_j}{n_{\varepsilon_j}}
\sqrt{\frac{\varepsilon_0}{\mu_0}}
\sin(k_{\rm THz}n_{\varepsilon_j}\Delta z),
\\
&
\mathbb{M}_{j,21}
=
i\frac{n_{\varepsilon_j}}{\varepsilon_j}
\sqrt{\frac{\mu_0}{\varepsilon_0}}
\sin(k_{\rm THz}n_{\varepsilon_j} \Delta z),
\end{align}
where $k_{\rm THz}$ is the free-space wavevector of the sub-THz fields, and $\theta_{\rm i}$ is taken as the Brewster's angle $\theta_{\rm B}$. Using Eq.~(\ref{EQ:connection_tb}), we build up the relation between the sub-THz fields on the top and bottom surfaces of the Si wafer as
\begin{align}
&\begin{pmatrix}
U(0)\\
V(0)
\end{pmatrix}
=
(\mathbb{M}_{q+1}
\mathbb{M}_{q}
...
\mathbb{M}_{1})^{-1}
\begin{pmatrix}
U(L_{\rm Si})\\
V(L_{\rm Si})
\end{pmatrix}\notag\\
\equiv&
\mathcal{M}
\begin{pmatrix}
U(L_{\rm Si})\\
V(L_{\rm Si})
\end{pmatrix}
=
\begin{pmatrix}
\mathcal{M}_{11} & \mathcal{M}_{12}\\
\mathcal{M}_{21} & \mathcal{M}_{22}
\end{pmatrix}
\begin{pmatrix}
U(L_{\rm Si})\\
V(L_{\rm Si})
\end{pmatrix}.
\label{EQ:total_transfermat}
\end{align}
Similar to Eqs.~(\ref{EQ:gr_u0}),~(\ref{EQ:gr_v0}),~(\ref{EQ:gr_u}), and ~(\ref{EQ:gr_vw}) in the problem of the green-light progapation, we can write
\begin{align}
&U(0)
=
H_{\rm i}+H_{\rm r},
\\
&V(0)
=
\frac{\cos\theta_{\rm B}}{n_{\rm Air}\sqrt{\varepsilon_0/\mu_0}}
(H_{\rm i}-H_{\rm r}),
\\
&U(L_{\rm Si})
=
H_{\rm t},
\\
&V(L_{\rm Si})
=
\frac{\cos\theta_{\rm B}}{n_{\rm Air}\sqrt{\varepsilon_0/\mu_0}}
H_{\rm t},
\end{align}
where the parameters $H_{\rm i}$, $H_{\rm r}$, and $H_{\rm t}$ define the reflection and transmission coefficients, $r_{\rm THz}=H_{\rm r}/H_{\rm i}$ and $t_{\rm THz}=H_{\rm t}/H_{\rm i}$. Taking the ratio $V(0)/U(0)$ and using Eq.~(\ref{EQ:total_transfermat}), we obtain an equation for the reflection coefficient $r_{\rm THz}$,
\begin{align}
&\frac{\cos\theta_{\rm B}}{n_{\rm Air}\sqrt{\varepsilon_0/\mu_0}}\frac{1-r_{\rm THz}}{1+r_{\rm THz}}
\notag\\
=
&
\frac{\mathcal{M}_{21} + \frac{\cos\theta_{\rm B}}{n_{\rm Air}\sqrt{\varepsilon_0/\mu_0}}\mathcal{M}_{22}}{\mathcal{M}_{11} + \frac{\cos\theta_{\rm B}}{n_{\rm Air}\sqrt{\varepsilon_0/\mu_0}}\mathcal{M}_{12}},
\end{align}
which can be easily solved as
\begin{widetext}
\begin{align}
r_{\rm THz}
=
\frac
{
\mathcal{M}_{11} 
-\mathcal{M}_{22}
+ \frac{\cos\theta_{\rm B}}{n_{\rm Air}\sqrt{\varepsilon_0/\mu_0}}\mathcal{M}_{12}
-(\frac{\cos\theta_{\rm B}}{n_{\rm Air}\sqrt{\varepsilon_0/\mu_0}})^{-1}\mathcal{M}_{21}
}
{
\mathcal{M}_{11} 
+\mathcal{M}_{22}
+ \frac{\cos\theta_{\rm B}}{n_{\rm Air}\sqrt{\varepsilon_0/\mu_0}}\mathcal{M}_{12}
+(\frac{\cos\theta_{\rm B}}{n_{\rm Air}\sqrt{\varepsilon_0/\mu_0}})^{-1}\mathcal{M}_{21}
}.
\end{align}
\end{widetext}
The reflectance for the sub-THz waves is then calculated as $|r_{\rm THz}|^2$.

%

\end{document}